\documentclass[preprint,twocolumn,10pt,notitlepage,nofootinbib,pra]{revtex4}%
\usepackage{amsfonts}
\usepackage{amsmath}
\usepackage{amssymb}
\usepackage{graphicx}%
\begin{document}
\title{Semiclassical Theory of Amplification and Lasing }
\author{Germ\'{a}n J. de Valc\'{a}rcel and Eugenio Rold\'{a}n}
\affiliation{Departament d'\`{O}ptica, Universitat de Val\`{e}ncia, Dr. Moliner 50,
46100--Burjassot, Spain}
\author{F. Prati}
\affiliation{CNISM and Dipartimento di Fisica e Matematica,
Universit\`a dell'Insubria, via Valleggio 11, 22100 Como, Italy}

\begin{abstract}
In this article we present a systematic derivation of the Maxwell--Bloch
equations describing amplification and laser action in a ring cavity. We
derive the Maxwell--Bloch equations for a two--level medium and discuss their
applicability to standard three-- and four--level systems. After discusing
amplification, we consider lasing and pay special attention to the obtention
of the laser equations in the uniform field approximation. Finally, the
connection of the laser equations with the Lorenz model is considered.

\end{abstract}

\pacs{42.55.-f; 42.55.Ah; 42.50.-p}
\maketitle

\section{Introduction}

Laser theory is a major branch of quantum optics and there are many
textbooks devoted to that matter or that pay a special attention to
it (see, e.g.,
\cite{Sargent,Haken,Siegman,Yariv,MilonniEberly,Svelto,NarducciAbraham,Meystre,WeissVilaseca,Khanin,Silfvast,Mandel}).
In spite of this we do think that there is room for new didactic
presentations of the basic semiclassical laser theory equations as
some aspects are not properly covered in the standard didactic
material or are scattered in specialized sources. The most clear
example concerns the uniform field limit approximation
\cite{citaUFL}, which is usually assumed \textit{ab initio} without
discussion, and when discussed, as e.g. in \cite{NarducciAbraham},
it is done in a way that admits relevant simplifications. In fact,
this important approximation has found a correct form only recently
\cite{deValcarcel03}. Other important aspect that is usually missed
in textbooks is the applicability of the standard two--level
approximation to the more realistic three-- and four--level schemes.
Certainly this matter is discussed in some detail in \cite{Khanin}
but we find it important to insist on this as it is usually missed
and may lead to some misconceptions, as we discuss below.

There are many good general textbooks on the fundamentals of lasers, e.g.
\cite{Siegman,Yariv,Svelto,Silfvast}, and we refer the reader to any of them
for getting an overview on the general characteristics of the different laser
types. Here it will suffice to say a few words on the structure of the laser.
\begin{figure}[t]
\begin{center}
\scalebox{0.6}{\includegraphics{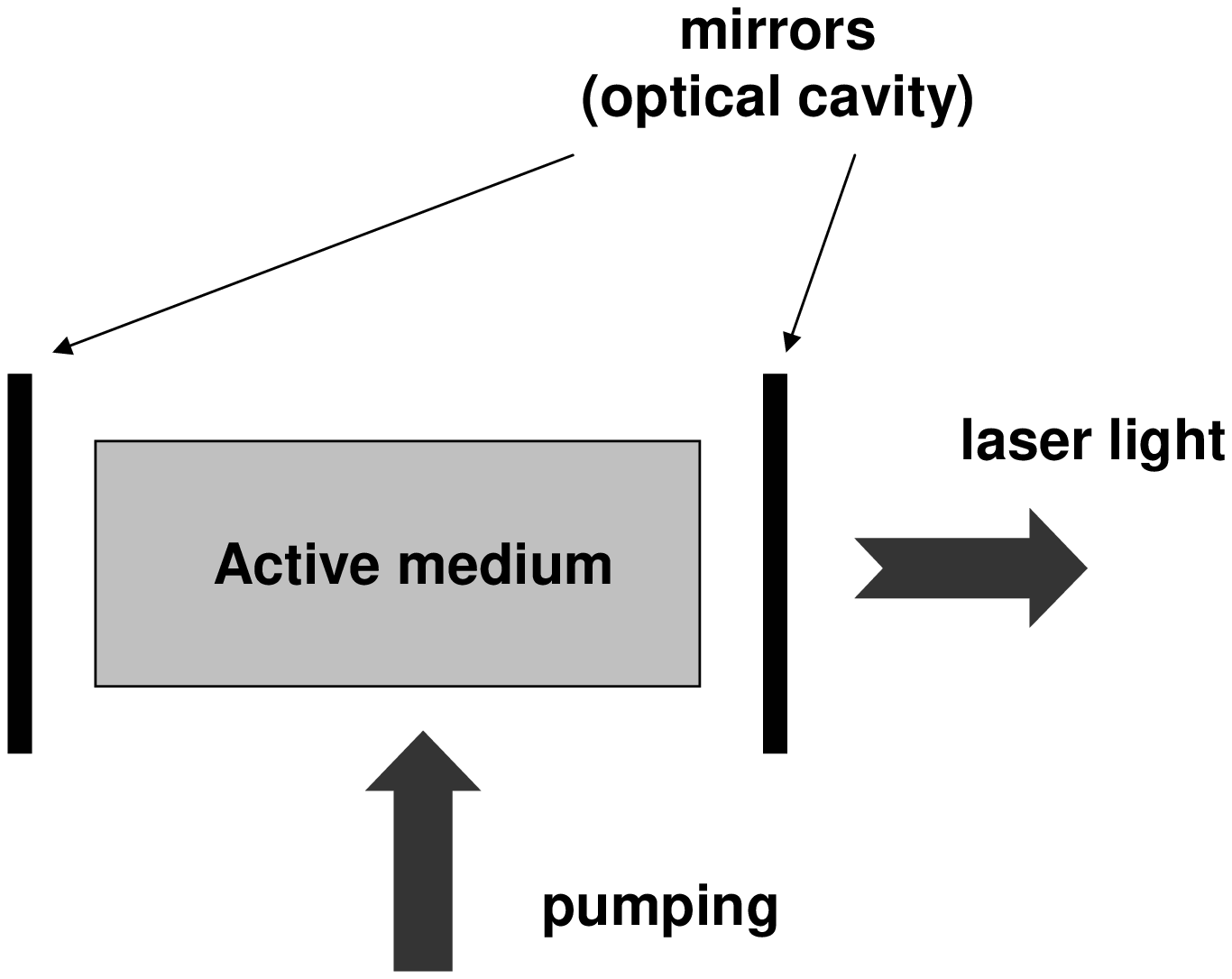}}
\end{center}
\caption{Scheme of a typical laser in a Fabry--Perot cavity.}
\end{figure}
A typical laser consists of three basic elements: An optical cavity,
an amplifying medium, and a pumping mechanism, see Fig. 1. The
optical cavity (also named resonator or oscillator) consists of two
or more mirrors that force light to propagate in a closed circuit,
further imposing it a certain modal structure. There are two basic
types of optical cavities, namely ring and linear, that differ in
the boundary condition that the cavity mirrors impose to the
intracavity field. In ring resonators the field inside the cavity
can be described as a traveling wave \footnote{The traveling wave
will propagate with a given sense of rotation, say, clockwise.
However, one could in principle expect a second field propagating
counter--clockwise, i.e., one could expect bidirectional emission in
a ring laser. Nevertheless, this is usually (although not always)
avoided by using some intracavity elements such as, e.g., Faraday
isolators. In any case, we shall not deal here with bidirectional
emission.}. Contrarily, in linear (also named Fabry--Perot--type)
resonators, the field is better described as a standing wave, which
requires a more complicated mathematical description than the
traveling wave case.

The amplifying medium can be solid, liquid, gas, or plasma. Nevertheless, most
cases are well described by considering that the amplifying medium consists of
a number of atoms, ions or molecules of which a number of states (energy
levels), with suitable relaxation rates and dipolar momenta, are involved in
the interaction with the electromagnetic field. It is customary to adopt the
so--called two--level approximation, i.e., to assume that only two energy
levels of the amplifying medium are relevant for the interaction. Actually a
minimum of three or four levels are necessary in order to obtain population
inversion, and we discuss below how the two--level theory applies to these
more complicated level schemes.

Then there is the pumping mechanism. This is highly specific for
each laser type but it has always the same purpose: Creating enough
population inversion for laser action. When modeling
radiation--matter interaction inside the laser cavity one can
usually forget the specifics of the pumping mechanism (whether it is
an electric current or a broadband optical discharge or whatever)
and describe it through a suitable pumping parameter. In this point,
the consideration of two--, three-- or four--level atomic schemes
turns out to be important, as it is here where the pumping mechanism
affects the mathematical description as we show below.

Laser physics studies all of these aspects of lasers but here we shall not
deal but with the mathematical description of the interaction between light
and matter inside the laser cavity. In this article we shall provide a
systematic derivation of the semiclassical laser equations for an important
and simple case:\ The homogeneously broadened ring laser, which plays the role
of a paradigm in laser physics. We shall not consider the important issues of
inhomogeneous broadening or linear optical resonators, because we want to
maintain the derivation as simple as possible (but not more!).

As stated, we shall use semiclassical theory, i.e., we shall consider a
classical electromagnetic field in interaction with a quantized medium. The
quantization of the medium is necessary in order to correctly describe
absorption and amplification as the classical theory (that models matter as a
collection of forced and damped harmonic oscillators) cannot be used for that.
With respect to the quantization of the field, it is not necessary if one (i)
is not interested in the field fluctuations, and (ii) accepts a heuristic
description of relaxation phenomena (in particular of spontaneous emission).
In any case, the quantum theory of the laser requires the use of complicated
mathematical techniques and falls outside the scope of our interests here.

After this introduction the rest of the article is organized as follows: In
Section II we derive the field equation; in Section III we derive the matter
equations for two--, three--, and four--level atoms or molecules; and in
Section IV we connect these with the field equation and write down the
Maxwell--Bloch equations. Then Sections V and VI are devoted to the analysis
of amplification and lasing, respectively. In Section VII we present a clear
derivation of the uniform field equations, and in Section VIII we present the
"Lorenz" from of the laser equations. Finally, in Section IX we present our conclusions.

\section{The Field Equation}

Maxwell's equations for a nonmagnetic material without free charges yield the
wave equation
\begin{equation}
\boldsymbol{\nabla}^{2}\mathbf{E}-c^{-2}\partial_{t}^{2}\mathbf{E}%
+\boldsymbol{\nabla}\left(  \boldsymbol{\nabla}\cdot\mathbf{E}\right)
=\mu_{0}\partial_{t}^{2}\mathbf{P}. \label{MWE}%
\end{equation}
Along this article we shall assume that the electric field $\mathbf{E}$ is a
plane wave propagating along the $z$ axis, and write it in the form%
\begin{equation}
\mathbf{E}\left(  \mathbf{r},t\right)  =\tfrac{1}{2}\mathbf{e}\mathcal{E}%
\left(  z,t\right)  e^{i\left(  kz-\omega t\right)  }+c.c., \label{E}%
\end{equation}
where $\mathbf{e}$ is the unit polarization vector (fixed polarization is
assumed), and%
\begin{equation}
k=\omega/c. \label{dispersion0}%
\end{equation}
We note that $\omega$ is an arbitrary reference (carrier) frequency. For
instance, if light is perfectly monochromatic of frequency $\omega_{0}$ we can
still choose $\omega\neq\omega_{0}$ as we allow the complex amplitude
$\mathcal{E}\left(  z,t\right)  $ to be time and space dependent. The
situation is even clearer when dealing with light whose spectrum has some
finite width: In this case even the concept of "light frequency" is
ill-defined, and clearly $\omega$ can be chosen arbitrarily.

Given the form (\ref{E}) for the electric field, by consistency with the wave
equation, the polarization $\mathbf{P}$ must read%
\begin{equation}
\mathbf{P}\left(  \mathbf{r},t\right)  =\tfrac{1}{2}\mathbf{e}\mathcal{P}%
\left(  z,t\right)  e^{i\left(  kz-\omega t\right)  }+c.c.. \label{P}%
\end{equation}

Now one must substitute these expressions into the wave equation and perform
the \textit{Slowly Varying Envelope Approximation} (SVEA) that consists in
assuming that%
\begin{align}
\partial_{t}^{2}U  &  \ll\omega\partial_{t}U\ll\omega^{2}U,\\
\partial_{z}^{2}U  &  \ll k\partial_{z}U\ll k^{2}U,
\end{align}
for $U=\mathcal{E}$ or $\mathcal{P}$. The physical meaning of this important
approximation is clear: One considers that temporal (spatial) variations of
the amplitudes $U$ contain temporal (spatial) frequencies that are much
smaller than the carrier frequency (wavenumber). In other words: The
amplitudes $U$ are assumed to vary on time (space) scales much slower (longer)
than the optical frequency (wavelength). Obviously this approximation excludes
the (limit) case of ultrashort pulses containing only a few cycles of the
field, but it is overall very accurate in general, even for short pulses as
soon as a sufficient number of cycles enter within the pulse width.

After performing the SVEA and multiplying the resulting equation by
$c^{2}/2i\omega$, one readily obtains%
\begin{equation}
\left(  \partial_{t}+c\partial_{z}\right)  \mathcal{E}=i\frac{\omega
}{2\varepsilon_{0}}\mathcal{P}, \label{maxE}%
\end{equation}
which is the field equation of interest. Let us remark that the SVEA is a
fundamental approximation in laser theory as it allows to transform the
original wave equation, which is a second-order partial differential equation
(PDE), into a first-order PDE.

Now we need to calculate the source term $\mathcal{P}$ and we do this in the
next section.

\section{The matter equations: optical Bloch equations}

The wave equation (\ref{maxE}) relates the slowly varying electric field
amplitude $\mathcal{E}$ with its source, the slowly varying polarization
amplitude $\mathcal{P}$. We discuss in this section how this last quantity is
determined. First we introduce the two--level atom model and derive the
evolution equation for its density matrix, the so-called optical Bloch
equations. Next, the density matrix is shown to yield the information
necessary for computing $\mathcal{P}$ what allows to write a closed set of
equations describing the coupled evolution of field and matter, the
Maxwell--Bloch equations. Then we consider the case of three-- and four--level
atoms, which is a more realistic approximation to actual lasers. After
deriving their corresponding Bloch equations, we discuss the conditions under
which the two--level model can be applied to three-- and four--level atoms. In
particular this is a necessary step for correctly understand the meaning of
the pump parameter.

\subsection{The two--level atom model}

\subsubsection{Hamiltonian}

The Hamiltonian of the system consists of two pieces: One describing the atom
or molecule in the absence of electromagnetic interaction, and the other
describing the action of the electromagnetic field on this atom, i.e.%
\begin{equation}
H\left(  \mathbf{r},t\right)  =H_{\mathrm{at}}+H_{\mathrm{int}}\left(
\mathbf{r},t\right)  . \label{ham}%
\end{equation}
\begin{figure}[t]
\begin{center}
\scalebox{0.8}{\includegraphics{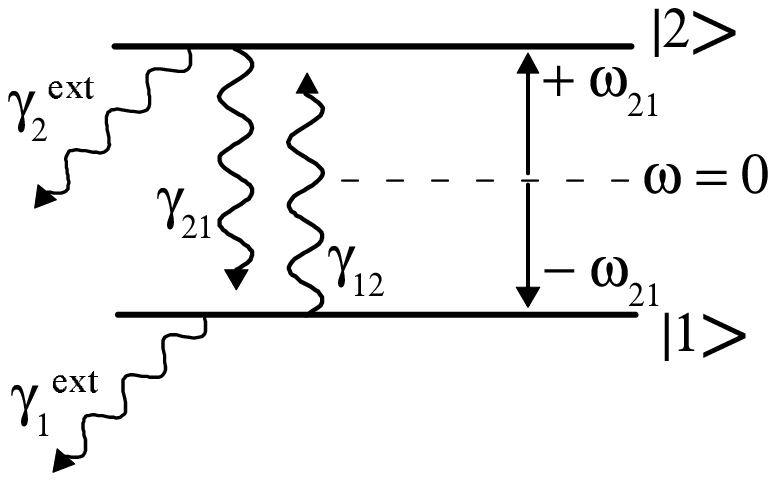}}
\end{center}
\caption{Schematic of the two--level atom energy levels including
relaxation rates (see text)}
\end{figure}
The material medium is assumed to be a system of identical two--level atoms or
molecules, i.e., it is assumed that the material medium is homogeneously
broadened. We denote by $\left\vert 1\right\rangle $ and $\left\vert
2\right\rangle $ the lower and higher energy levels, respectively, and by
$\omega_{21}$ the transition frequency of one of these atoms. This means that
the atomic Hamiltonian $H_{\mathrm{at}}$ verifies%

\begin{align}
H_{\mathrm{at}}\left\vert 2\right\rangle  &  =+\tfrac{1}{2}\hbar\omega
_{21}\left\vert 2\right\rangle ,\\
H_{\mathrm{at}}\left\vert 1\right\rangle  &  =-\tfrac{1}{2}\hbar\omega
_{21}\left\vert 1\right\rangle ,
\end{align}
where we have chosen the arbitrary (and unimportant) energy origin in such a
way that it lies halfway between both states energies (see, Fig. 2). The
matrix representation for this Hamiltonian thus reads%
\begin{equation}
H_{\mathrm{at}}=%
\begin{bmatrix}
+\tfrac{1}{2}\hbar\omega_{21} & 0\\
0 & -\tfrac{1}{2}\hbar\omega_{21}%
\end{bmatrix}
, \label{Hat}%
\end{equation}
where the level ordering has been chosen to be $\left\{  \left\vert
2\right\rangle ,\left\vert 1\right\rangle \right\}  $.

The interaction Hamiltonian $H_{\mathrm{int}}$ is taken in the electric dipole
approximation. Roughly speaking, this approximation is valid when the light
wavelength is much longer than the typical dimensions of the electronic cloud,
which is on the order of $1$ $%
\operatorname{\text{\AA}}%
$. Thus the approximation is justified in the infrared and visible parts of
the spectrum and even in the ultraviolet. This interaction hamiltonian reads%
\begin{equation}
H_{\mathrm{int}}\left(  \mathbf{r},t\right)  =-\boldsymbol{\hat{\mu}}%
\cdot\mathbf{E}\left(  \mathbf{r},t\right)  ,
\end{equation}
where $\mathbf{r}$ denotes the position of the atom (which is not quantized in
the theory) and the operator $\boldsymbol{\hat{\mu}}=-e\mathbf{\hat{r}%
}_{\mathrm{at}}$, being $-e$ the electron charge and $\mathbf{\hat{r}%
}_{\mathrm{at}}$ the vector position operator of the electron relative to the
point-like nucleus. $\boldsymbol{\hat{\mu}}$ acts on the atomic variables
whereas in this semiclassical formalism the field $\mathbf{E}$ is a
\textrm{c}-number. In the chosen basis ordering the matrix form for this
hamiltonian reads%
\begin{equation}
H_{\mathrm{int}}\left(  \mathbf{r},t\right)  =%
\begin{bmatrix}
-\boldsymbol{\mu}_{22}\cdot\mathbf{E}\left(  \mathbf{r},t\right)  \ \  &
-\boldsymbol{\mu}_{21}\cdot\mathbf{E}\left(  \mathbf{r},t\right) \\
-\boldsymbol{\mu}_{12}\cdot\mathbf{E}\left(  \mathbf{r},t\right)  \ \  &
-\boldsymbol{\mu}_{11}\cdot\mathbf{E}\left(  \mathbf{r},t\right)
\end{bmatrix}
, \label{Hint1}%
\end{equation}
where the matrix elements%
\begin{equation}
\boldsymbol{\mu}_{mn}=\left\langle m\right\vert \boldsymbol{\hat{\mu}%
}\left\vert n\right\rangle \equiv-e\int d^{3}r_{\mathrm{at}}\ \psi_{m}^{\ast
}\left(  \mathbf{\hat{r}}_{\mathrm{at}}\right)  \mathbf{\hat{r}}_{\mathrm{at}%
}\psi_{n}\left(  \mathbf{\hat{r}}_{\mathrm{at}}\right)  ,
\end{equation}
and $\psi_{n}\left(  \mathbf{\hat{r}}_{\mathrm{at}}\right)  $ is the
wavefunction (in position representation) of the atomic state $\left\vert
n\right\rangle $. (Note that $\boldsymbol{\mu}_{mn}=\boldsymbol{\mu}%
_{nm}^{\ast}$.) We now recall the parity property of atomic eigenstates: All
atomic eigenstates have well defined parity (even or odd) due to the central
character of the atomic potential. This means that $\boldsymbol{\mu}%
_{11}=\boldsymbol{\mu}_{22}=0$ and then, in order to have interaction, we must
consider states $\left\vert 1\right\rangle $ and $\left\vert 2\right\rangle $
with opposite parity (this is the basic selection rule of atomic transitions
in the electric dipole approximation). Hence the interaction hamiltonian
(\ref{Hint1}) becomes%
\begin{equation}
H_{\mathrm{int}}\left(  \mathbf{r},t\right)  =%
\begin{bmatrix}
0 & V\left(  \mathbf{r},t\right) \\
V^{\ast}\left(  \mathbf{r},t\right)  & 0
\end{bmatrix}
, \label{Hint}%
\end{equation}
where we have introduced the notation%
\begin{equation}
V\left(  \mathbf{r},t\right)  =-\boldsymbol{\mu}_{21}\cdot\mathbf{E}\left(
\mathbf{r},t\right)  . \label{V}%
\end{equation}
Taking into account the form of the electric field, Eq. (\ref{E}), $V\left(
\mathbf{r},t\right)  $ becomes%
\begin{equation}
V\left(  \mathbf{r},t\right)  =-\hbar\alpha\left(  z,t\right)  e^{i\left(
kz-\omega t\right)  }-\hbar\beta\left(  z,t\right)  e^{-i\left(  kz-\omega
t\right)  }, \label{Va}%
\end{equation}
where we have defined%
\begin{equation}
\alpha\left(  z,t\right)  =\frac{\boldsymbol{\mu}_{21}\cdot\mathbf{e}}{2\hbar
}\mathcal{E}\left(  z,t\right)  ,\qquad\beta\left(  z,t\right)  =\frac
{\boldsymbol{\mu}_{21}\cdot\mathbf{e}^{\ast}}{2\hbar}\mathcal{E}^{\ast}\left(
z,t\right)  . \label{alpha}%
\end{equation}
We note that $2\alpha$ is usually referred to as the (complex) Rabi frequency
of the light field.

Finally the total hamiltonian $H$, Eq. (\ref{ham}), for a two--level atom
located at position $\mathbf{r}$ interacting with a light field reads%
\begin{equation}
H\left(  \mathbf{r},t\right)  =%
\begin{bmatrix}
+\tfrac{1}{2}\hbar\omega_{21}\ \  & V\left(  \mathbf{r},t\right) \\
V^{\ast}\left(  \mathbf{r},t\right)  \ \  & -\tfrac{1}{2}\hbar\omega_{21}%
\end{bmatrix}
. \label{H}%
\end{equation}

\subsubsection{The density matrix. Evolution}

The hamiltonian $H$ can serve us to write down the Schr\"{o}dinger equation
for the atomic wavefunction. Instead, we use here the density matrix formalism
as it is the most appropriate in order to incorporate damping and pumping
terms into the equations of motion, something we shall do in the next
subsection. In the chosen basis ordering, the density matrix $\rho$
representing a two--level atom located at $\mathbf{r}$ takes the form%
\begin{equation}
\rho\left(  \mathbf{r},t\right)  =%
\begin{bmatrix}
\rho_{22}\left(  \mathbf{r},t\right)  \ \  & \rho_{21}\left(  \mathbf{r}%
,t\right) \\
\rho_{12}\left(  \mathbf{r},t\right)  \ \  & \rho_{11}\left(  \mathbf{r}%
,t\right)
\end{bmatrix}
. \label{rho}%
\end{equation}

The meaning of the matrix elements is as follows: $\rho_{mm}$ denotes the
probability ($0\leq\rho_{mm}\leq1$) that the atom occupies state $\left\vert
m\right\rangle $, and $\rho_{mn}\left(  =\rho_{nm}^{\ast}\right)  $ is the
coherence between the two atomic states, which is related with the
polarization induced in the the atom by the light field; see below. The
evolution of $\rho$ is governed by the Schr\"{o}dinger--von Neumann equation%
\begin{equation}
i\hbar\partial_{t}\rho=\left[  H,\rho\right]  . \label{von}%
\end{equation}
Upon substituting Eqs. (\ref{rho}) and (\ref{H}) into Eq. (\ref{von}) one
obtains a set of equations which is simplified by defining the new variables%
\begin{equation}
\sigma_{12}=\sigma_{21}^{\ast}=\rho_{12}e^{i\left(  kz-\omega t\right)  }.
\label{s12-s21}%
\end{equation}
This is motivated by the functional dependence of the nondiagonal elements
$\rho_{12}$ and $\rho_{21}$ on space and time under free evolution ($V=0$).
(We note that the above transformation is equivalent to working in the
so-called interaction picture of quantum mechanics.) The explicit space-time
dependence added in Eq. (\ref{s12-s21}) makes that the new quantities
$\sigma_{ij}$ are slowly varying as will become evident later. In terms of
these reduced density matrix elements, and making use of Eq. (\ref{Va}), the
Schr\"{o}dinger--von Neumann equation (\ref{von}) becomes%
\begin{align}
\partial_{t}\rho_{22}  &  =i\alpha\sigma_{12}+i\beta\sigma_{12}e^{-2i\left(
kz-\omega t\right)  }+c.c.,\label{e1a}\\
\partial_{t}\rho_{11}  &  =-i\alpha\sigma_{12}-i\beta\sigma_{12}e^{-2i\left(
kz-\omega t\right)  }+c.c.,\label{e2a}\\
\partial_{t}\sigma_{12}  &  =-i\delta\sigma_{12}+i\left(  \rho_{22}-\rho
_{11}\right)  \left[  \alpha^{\ast}+\beta^{\ast}e^{2i\left(  kz-\omega
t\right)  }\right]  , \label{e3a}%
\end{align}
where we have introduced the mistuning, or detuning, parameter%
\begin{equation}
\delta=\omega-\omega_{21}. \label{delta}%
\end{equation}
Note that $\partial_{t}\left(  \rho_{22}+\rho_{11}\right)  =0$, what implies
probability conservation.

We now make a most important and widely used approximation in quantum optics,
namely the \textit{Rotating Wave Approximation} (RWA). An inspection of Eqs.
(\ref{e1a})--(\ref{e3a}) shows that, in the absence of interaction
($\mathcal{E}=0$, i.e. $\alpha=\beta=0$ in the new notation), $\rho_{22}$ and
$\rho_{11}$ are constant and $\sigma_{12}=\sigma_{21}^{\ast}$ oscillate at the
low (non optical) frequency $\delta$. This means that the time scales of the
free system are large as compared with the optical periods. Now, if the
interaction is turned on we see in Eqs. (\ref{e1a})--(\ref{e3a}) that slowly
varying terms (those proportional to $\alpha$ or $\alpha^{\ast}$) appear, as
well as high frequency terms oscillating as $\exp\left[  \pm2i\left(
kz-\omega t\right)  \right]  $ (the terms proportional to $\beta$ or
$\beta^{\ast}$). Clearly the atom cannot respond to the latter and one can
discard them. This is the RWA, which can be easily demonstrated by using
perturbation theory.

After performing the RWA, Eqs. (\ref{e1a})--(\ref{e3a}) become%
\begin{align}
\partial_{t}\rho_{22}  &  =i\left(  \alpha\sigma_{12}-\alpha^{\ast}\sigma
_{21}\right)  ,\label{Bloch1}\\
\partial_{t}\rho_{11}  &  =-i\left(  \alpha\sigma_{12}-\alpha^{\ast}%
\sigma_{21}\right)  ,\label{Bloch2}\\
\partial_{t}\sigma_{12}  &  =-i\delta\sigma_{12}+i\alpha^{\ast}\left(
\rho_{22}-\rho_{11}\right)  , \label{Bloch3}%
\end{align}
which is the standard form of the optical Bloch equations for a single atom.

\subsubsection{The population matrix}

We are dealing with a situation in which there is not a single atom or
molecule interacting with the light field but a very large number of them, and
then some ensemble averaging must be performed. The ensemble averaged density
matrix is called the \textit{population matrix} \cite{Sargent}, although the
name density matrix is more frequently used obscuring the differences between
the two operators. Here we are not going to introduce the population matrix
rigorously and we refer the interested reader to \cite{Sargent} or
\cite{Meystre} for further details.

The population matrix of an ensemble of molecules is defined as
\begin{equation}
\bar{\rho}\left(  z,t\right)  =\mathcal{N}^{-1}%
{\textstyle\sum_{a}}
\rho_{a}\left(  z,t\right)  . \label{population}%
\end{equation}
Here\footnote{Here we are considering a plane wave laser beam propagating
along $z$; hence atoms are grouped according to that coordinate. In the
general three--dimensional case, a population matrix $\bar{\rho}\left(
\mathbf{r},t\right)  $ must be defined at every differential volume,
analogously to (\ref{population}).} $\bar{\rho}$ is the population matrix,
$\rho_{a}$ is the density matrix for an atom labeled by $a$, and $a$ runs
along all molecules that, at time $t$, are within $z$ and $z+dz$ .
$\mathcal{N}$ is the number of such molecules, which is assumed to be
independent of $z$ and $t$. The equation of evolution of the population matrix
has two contributions: One of them is formally like the Schr\"{o}dinger--von
Neumann equation governing the evolution of the density matrix of a single
atom, and the other one describes incoherent processes (i.e. not due to the
interaction with the electromagnetic field such as pumping and relaxation
phenomena due to collisions between atoms or to spontaneous emission)
\cite{Sargent}
\begin{equation}
\partial_{t}\bar{\rho}_{ij}=\left(  i\hbar\right)  ^{-1}\left[  H,\bar{\rho
}\right]  _{ij}+\left(  \hat{\Gamma}\bar{\rho}\right)  _{ij},
\label{vonneumann}%
\end{equation}
$\left(  i,j=1,2\right)  $. In Eq. (\ref{vonneumann}) the term $\hat{\Gamma
}\bar{\rho}$ is the one describing incoherent processes and $\hat{\Gamma}$ is
the Liouville (super)operator.

Consider the situation depicted in Fig. 2. It corresponds to the following
matrix elements for the operator $\left(  \hat{\Gamma}\bar{\rho}\right)  $
\begin{align}
\left(  \hat{\Gamma}\bar{\rho}\right)  _{22}  &  =-\gamma_{2}\bar{\rho}%
_{22}+\gamma_{12}\bar{\rho}_{11}+\lambda_{2},\nonumber\\
\left(  \hat{\Gamma}\bar{\rho}\right)  _{21}  &  =\left(  \hat{\Gamma}%
\bar{\rho}\right)  _{12}^{\ast}=-\gamma_{\bot}\bar{\rho}_{21},\\
\left(  \hat{\Gamma}\bar{\rho}\right)  _{11}  &  =-\gamma_{1}\bar{\rho}%
_{11}+\gamma_{21}\bar{\rho}_{22}+\lambda_{1},\nonumber
\end{align}
where%
\begin{align}
\gamma_{2}  &  =\gamma_{2}^{ext}+\gamma_{21},\\
\gamma_{1}  &  =\gamma_{1}^{ext}+\gamma_{12}.
\end{align}
In the above expressions, $\gamma_{ij}$ describes the relaxation rate from
level $\left\vert i\right\rangle $ to level $\left\vert j\right\rangle $ (that
is, the pass of population from level $\left\vert i\right\rangle $ to level
$\left\vert j\right\rangle $ due to collisions), and $\gamma_{i}^{ext}$ the
relaxation rate from level $\left\vert i\right\rangle $ to some other external
level (see Fig. 2). The term $\lambda_{i}$ is the pumping rate of level
$\left\vert i\right\rangle $, i.e., it describes the increase of population of
level $\left\vert i\right\rangle $ due to pumping processes. Notice that it is
not specified from where this population is coming as only the dynamics of the
two lasing level populations is being described. We shall come back to this
important point in the following subsection.

The value of the different decay constants appearing in $\hat{\Gamma}\bar
{\rho}$ depend strongly on the particular substance and operating conditions.
In any case it is always verified that
\begin{equation}
\gamma_{\bot}\geq\frac{1}{2}\left(  \gamma_{2}+\gamma_{1}\right)  ,
\end{equation}
what reflects the fact that the coherence $\rho_{ij}$ is affected not only by
the relaxation mechanisms affecting the populations, but also by some specific
collisions, known as dephasing collisions, which do not affect the populations.

With the above form for the Liouvillian, the population matrix equations of
evolution read%
\begin{align}
\partial_{t}\rho_{22}  &  =-\gamma_{2}\rho_{22}+\gamma_{12}\rho_{11}%
+\lambda_{2}+i\left(  \alpha\sigma_{12}-\alpha^{\ast}\sigma_{21}\right)
,\label{Bloch4}\\
\partial_{t}\rho_{11}  &  =-\gamma_{1}\rho_{11}+\gamma_{21}\rho_{22}%
+\lambda_{1}-i\left(  \alpha\sigma_{12}-\alpha^{\ast}\sigma_{21}\right)
,\label{Bloch5}\\
\partial_{t}\sigma_{12}  &  =-\left(  \gamma_{\bot}+i\delta\right)
\sigma_{12}+i\alpha^{\ast}\left(  \rho_{22}-\rho_{11}\right)  , \label{Bloch6}%
\end{align}
where we have removed the overbar in order not to complicate unnecessarily the
notation. Now $\rho_{ii}$ can be understood as the fraction of atoms occupying
level $\left\vert i\right\rangle $, i.e., it is the population of this level.
Notice that in Eqs. (\ref{Bloch4}-\ref{Bloch6}) $\partial_{t}\left(  \rho
_{22}+\rho_{11}\right)  \neq0$ in general, what reflects the fact that the
system formed by the atomic levels $\left\vert 2\right\rangle $ and
$\left\vert 1\right\rangle $ is an open system in which population is gained
and lost through incoherent processes.

In the two--level laser model, internal relaxation processes (those governed
by $\gamma_{21}$ and $\gamma_{12}$) are usually neglected, and it is further
assumed that the two lasing levels relax to the external reservoir with the
same rate $\gamma_{||}=\gamma_{2}^{ext}=\gamma_{1}^{ext}$. It is easy to see
that in this simplified description of relaxation processes, the pumping
rates
\begin{equation}
\lambda_{i}=\frac{\rho_{ii}^{0}}{\gamma_{||}},
\end{equation}
with $\rho_{ii}^{0}$ the population of level $\left\vert i\right\rangle $ in
absence of fields ($\alpha=0$). Moreover, in this particular case
$\partial_{t}\left(  \rho_{22}+\rho_{11}\right)  =0$ and then a single
equation is needed for the description of the populations evolution. The
population difference is then defined%
\begin{equation}
d=\rho_{22}-\rho_{11},
\end{equation}
and Eqs. (\ref{Bloch4}-\ref{Bloch6}) simplify to%

\begin{align}
\partial_{t}d  &  =\gamma_{||}\left(  d_{0}-d\right)  +2i\left(  \alpha
\sigma_{12}-\alpha^{\ast}\sigma_{21}\right)  ,\label{B1}\\
\partial_{t}\sigma_{12}  &  =-\left(  \gamma_{\bot}+i\delta\right)
\sigma_{12}+i\alpha^{\ast}d, \label{B2}%
\end{align}
where
\begin{equation}
d_{0}=\rho_{22}^{0}-\rho_{11}^{0},
\end{equation}
is the population difference in the absence of fields, that is, the pump
parameter. This is the simplest way of modeling pumping. Clearly, $d_{0}>0$
implies an inverted medium (with a larger number of excited atoms than of
atoms in the fundamental state). If pumping is absent $d_{0}=-1$. Notice that
$d_{0}$ appears as a free parameter, that we can take positive or negative,
although we have not discussed yet how could it be controlled.

\subsubsection{Rate Equations}

It is interesting to write down Eqs. (\ref{Bloch4}-\ref{Bloch6}) when
$\gamma_{\bot}\gg\gamma_{||},\delta$ as in this case the adiabatic elimination
of the atomic polarization is justified (see Appendix 1). This adiabatic
elimination consists in making $\partial_{t}\sigma_{12}=0$, and then Eqs.
(\ref{Bloch4}-\ref{Bloch6}) reduce to%
\begin{align}
\partial_{t}\rho_{22}  &  =\lambda_{2}-\gamma_{2}\rho_{22}+\gamma_{12}%
\rho_{11}-R\left(  \rho_{22}-\rho_{11}\right)  ,\label{rate1}\\
\partial_{t}\rho_{11}  &  =\lambda_{1}-\gamma_{1}\rho_{11}+\gamma_{21}%
\rho_{22}+R\left(  \rho_{22}-\rho_{11}\right)  , \label{rate2}%
\end{align}
with $R=2\left\vert \alpha\right\vert ^{2}/\gamma_{\bot}.$

These equations are known as rate equations and are widely used in laser
physics as in most laser systems the condition for adiabatic elimination is
met. Let us remark that rate equations describe appropriately the interaction
of a light field with a two--level system in two limiting cases: When the
atomic polarization can be adiabatically eliminated, as we have discussed, and
also when the field is broadband (i.e., incoherent) in which case the factor
$R$ has a different expression to the one we have derived but again depends on
the square of the field amplitude \cite{deValcarcel04}. We shall make use of
these equations in the following subsection.

\subsection{Three--level and four--level atom models}

As we already commented in the introduction, actual lasers are based
on a three--level or four--level scheme rather than a two--level
one, the extra levels describing the reservoirs from which the pump
extracts atoms and to which damping sends atoms. In fact this extra
levels are necessary for obtaining population inversion ($d_{0}>0$)
which is a necessary condition for amplification and lasing, as we
show below. Although these extra levels do not participate directly
in laser action\footnote{There is a very important exception. In
coherent optically pumped lasers the pumping mechanism is a laser
field tuned to the pumping transition. If the atomic coherences
cannot be adiabatically eliminated, Raman processes are important
and cannot be neglected. We shall not consider these lasers here.},
the description of its indirect participation is essential in order
to correctly describe pumping and decaying processes. Here we shall
derive the Bloch equations for three--level and four--level atoms
interacting with a laser field and an incoherent pump, and connect
these equations with the two--level laser equations derived in the
previous section.

\subsubsection{Bloch equations for three--level atoms}
\begin{figure}[t]
\begin{center}
\scalebox{0.8}{\includegraphics{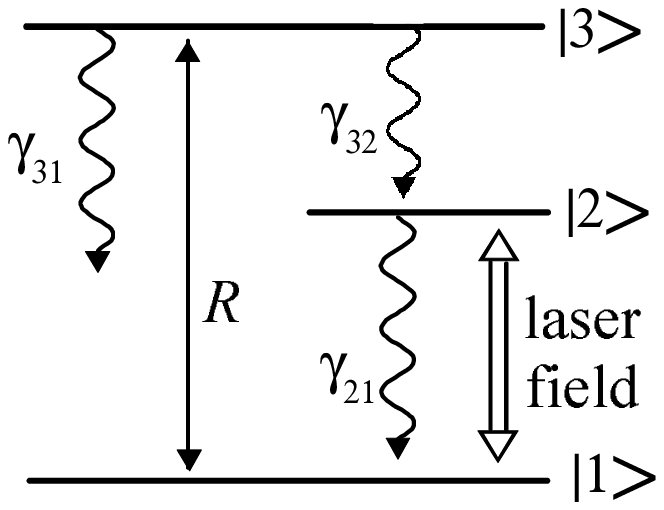}}
\end{center}
\caption{Schematic of a three--level atom. $R$ represents the
incoherent pumping and the laser field interacts with the
$\left\vert 2\right\rangle \rightarrow\left\vert 1\right\rangle $
transition. The arrows indicate decay processes.}
\end{figure}
Consider the three--level atom scheme depicted in Fig. 3, which can be
regarded as an approximate description of, e.g., the relevant atomic levels of
the Cr$^{3+}$or the Er$^{3+}$ ions that are the active ions in Ruby and Erbium
lasers, respectively. On these ions population is excited from the lower state
$\left\vert 1\right\rangle $ to the upper state $\left\vert 3\right\rangle $
by the pumping mechanism. Then population is transferred from level
$\left\vert 3\right\rangle $ to the upper lasing level $\left\vert
2\right\rangle $ (which is long--lived) by relaxation processes, which are
extremely fast in these ions.

We shall model the pumping transition $\left\vert 1\right\rangle
\longleftrightarrow\left\vert 3\right\rangle $ via rate equations
\footnote{For example, in the case of Ruby lasers, pumping comes
from an incoherent light source, namely a flashlamp. Then the
interaction of this incoherent light field with the pumping
transition $\left\vert 1\right\rangle \longleftrightarrow\left\vert
3\right\rangle $ can be described with the help of rate equations.
The case of Erbium lasers is different: In this case the pumping is
made with the help of a laser field tuned to the pumping transition.
In spite of the coherent nature of the pumping field, a rate
equations description for the pumping transition is also well suited
in this case, because the adiabatic elimination of the atomic
coherence of transition $\left\vert 1\right\rangle
\longleftrightarrow\left\vert 3\right\rangle $ is fully justified as
its coherence decay rate is very large as compared with the rest of
decay rates. Other systems have other pumping mechanisms (e.g., the
pass of an electrical current through the active medium) for which
the rate
equations description os also appropriate.} like Eqs. (\ref{rate1}%
,\ref{rate2}), and the interaction of the monochromatic field with transition
$\left\vert 1\right\rangle \longleftrightarrow\left\vert 2\right\rangle $ with
the already derived Bloch equations for a two--level atom. As for the
relaxation processes, we describe them heuristically, see Fig. 3. Then we can
model these processes with the following set of Bloch equations%
\begin{align}
\partial_{t}\rho_{33}  &  =-\left(  \gamma_{31}+\gamma_{32}\right)  \rho
_{33}+R\left(  \rho_{11}-\rho_{33}\right)  ,\label{three1}\\
\partial_{t}\rho_{22}  &  =-\gamma_{21}\rho_{22}+\gamma_{32}\rho_{33}+i\left(
\alpha\sigma_{12}-\alpha^{\ast}\sigma_{21}\right)  ,\label{three2}\\
\partial_{t}\rho_{11}  &  =\gamma_{21}\rho_{22}+\gamma_{31}\rho_{33}+R\left(
\rho_{33}-\rho_{11}\right) \label{three3}\\
&  -i\left(  \alpha\sigma_{12}-\alpha^{\ast}\sigma_{21}\right)  ,\nonumber\\
\partial_{t}\sigma_{12}  &  =-\left(  \gamma_{\bot}+i\delta\right)
\sigma_{21}+i\alpha^{\ast}\left(  \rho_{22}-\rho_{11}\right)  , \label{three4}%
\end{align}
where $R$ is the rate at which ions are pumped by the incoherent pump field
from level $\left\vert 1\right\rangle $ to level $\left\vert 3\right\rangle $.
Let us remark that in writing Eqs. (\ref{three1}) and (\ref{three3}): (i), we
have taken into account all possible transitions due to incoherent processes
with suitable relaxation rates as indicated in Fig. 3; and (ii), the
incoherent pumping of population from level $\left\vert 1\right\rangle $ to
level $\left\vert 3\right\rangle $ is modelled by the term $R\left(  \rho
_{11}-\rho_{33}\right)  $ appearing in Eqs. (\ref{three1}) and (\ref{three3})
with $R$ proportional to the pump intensity, i.e., we have described the
interaction of the pump field with the pumped transition by means of rate
equations similar to Eqs. (\ref{rate1}) and (\ref{rate2}) but taking
$\lambda_{i}=0$ as all incoherent processes have been consistently taken into account.

Let us further assume that $\gamma_{32}\gg\gamma_{21},\gamma_{31},\gamma
_{\bot},R$, as it occurs in usual three--level lasers. Then we adiabatically
eliminate the population of level $\left\vert 3\right\rangle $. By making
$\partial_{t}\rho_{33}=0$ we get%
\begin{equation}
\rho_{33}\approx\frac{R\rho_{11}}{\gamma_{32}}.
\end{equation}
This equation shows that $\gamma_{32}\rho_{33}$ is a finite quantity; that is,
$\rho_{33}$ is vanishingly small in the limit we are considering. Then we can
neglect $\rho_{33}$ in Eq. (\ref{three3}) and put $\gamma_{32}\rho_{33}%
=R\rho_{11}$ in Eq. (\ref{three2}). By further noticing that after the
approximation $\partial_{t}\left(  \rho_{11}+\rho_{22}\right)  =0$, we can
write the simplified model%
\begin{align}
\partial d  &  =R-\gamma_{21}-\left(  R+\gamma_{21}\right)  d+2i\left(
\alpha\sigma_{12}-\alpha^{\ast}\sigma_{21}\right)  ,\label{three5}\\
\partial_{t}\sigma_{12}  &  =-\left(  \gamma_{\bot}+i\delta\right)
\sigma_{12}+i\alpha^{\ast}d, \label{three6}%
\end{align}
where $d=\left(  \rho_{22}-\rho_{11}\right)  $. These are appropriate Bloch
equations for most three--level systems.
\begin{figure}[t]
\begin{center}
\scalebox{0.8}{\includegraphics{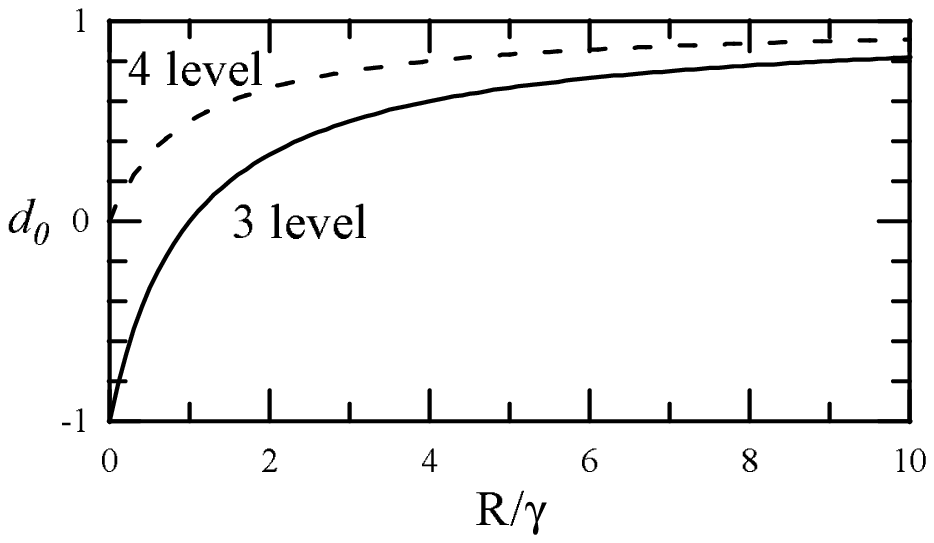}}
\end{center}
\caption{Dependence of the effective pump parameter $d_{0}$ on the
normalized actual pump strength $R/\gamma$ for three-- (full line)
and four--level (dashed line) lasers. $\gamma=\gamma_{21}$ for the
three--level laser and $\gamma=\gamma_{20}+\gamma_{21}$ for the
four--level laser.}
\end{figure}
We can now compare these equations that describe three--level atoms with Eqs.
(\ref{B1}) and (\ref{B2}) that describe two--level atoms in a simple and usual
limit. It is clear that they are isomorphic. Then we can conclude that
incoherently pumped three--level atoms can be described with the standard
two--level atom Bloch equations by making the following identifications%
\begin{align}
\gamma_{||}  &  \rightarrow R+\gamma_{21},\label{gammapar3L}\\
d_{0}  &  \rightarrow\frac{R-\gamma_{21}}{R+\gamma_{21}}. \label{pump3L}%
\end{align}
Notice that (i) the decay rate $\gamma_{||}$ is pump dependent for
three--level atoms, and (ii) that the pumping rate $d_{0}$ depends in a
nonlinear way with the actual pump parameter $R$. In Fig. 4 we represent
$d_{0}$ as a function of the actual pumping parameter $R$; notice that
increasing $R$ in a factor ten, say, does not mean to do so in $d_{0}$. Apart
from this, we have shown that an incoherently pumped three--level medium can
be described as a two--level one when the adiabatic eliminations we have
assumed are justified, which is the usual situation.

\subsubsection{Bloch equations for four--level atoms}
\begin{figure}[t]
\begin{center}
\scalebox{0.8}{\includegraphics{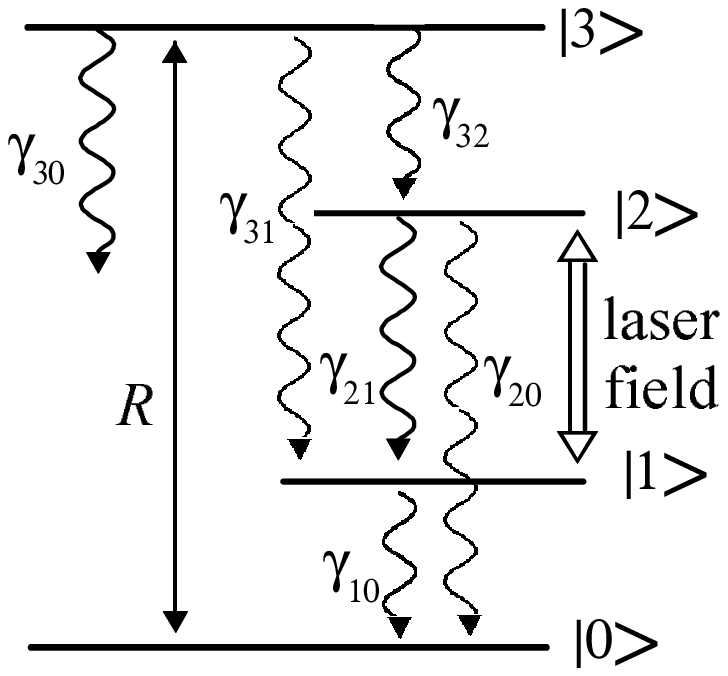}}
\end{center}
\caption{Schematic of a four--level atom. $R$ represents the
incoherent pumping and the laser field interacts with the
$\left\vert 2\right\rangle \rightarrow\left\vert 1\right\rangle $
transition. The arrows indicate decay processes.}
\end{figure}
Consider now the four--level atom scheme shown in Fig. 5. It can be regarded
as an approximate description of, e.g., the relevant atomic levels of the
$\mathrm{Nd}^{3+}$ ion that is the active ion Nd--YAG or Nd--glass lasers.
Assuming, as for three--level atoms, that the pumping field acting on the
transition $\left\vert 0\right\rangle -\left\vert 3\right\rangle $ can be
described by rate equations, we are left with the following optical Bloch
equations%
\begin{align}
\partial_{t}\rho_{33}  &  =-\left(  \gamma_{30}+\gamma_{31}+\gamma
_{32}\right)  \rho_{33}+R\left(  \rho_{00}-\rho_{33}\right)  ,\\
\partial_{t}\rho_{22}  &  =-\left(  \gamma_{20}+\gamma_{21}\right)  \rho
_{22}+\gamma_{32}\rho_{33}\\
&  +i\left(  \alpha\sigma_{12}-\alpha^{\ast}\sigma_{21}\right)  ,\nonumber\\
\partial_{t}\rho_{11}  &  =-\gamma_{10}\rho_{11}+\gamma_{21}\rho_{22}%
+\gamma_{31}\rho_{33}\\
&  +i\left(  \alpha\sigma_{12}-\alpha^{\ast}\sigma_{21}\right)  ,\nonumber\\
\partial_{t}\rho_{00}  &  =\gamma_{10}\rho_{11}+\gamma_{20}\rho_{22}%
+\gamma_{30}\rho_{33}-R\left(  \rho_{00}-\rho_{33}\right) \\
\partial_{t}\sigma_{12}  &  =-\left(  \gamma_{\bot}+i\delta\right)
\sigma_{21}+i\alpha^{\ast}\left(  \rho_{22}-\rho_{11}\right)  .
\end{align}
We can now proceed in a similar way as we did with three--level atoms: Let us
assume that $\gamma_{32}\ $is much larger than any other decay rate and
adiabatically eliminate $\rho_{33}$. Now we get%
\begin{equation}
\rho_{33}\approx\frac{R\rho_{00}}{\gamma_{32}},
\end{equation}
and neglecting the terms $\gamma_{31}\rho_{33},\gamma_{30}\rho_{33}$ and
$R\rho_{33}$ we are left with
\begin{align}
\partial_{t}\rho_{22}  &  =-\left(  \gamma_{20}+\gamma_{21}\right)  \rho
_{22}+R\rho_{00}\\
&  +i\left(  \alpha\sigma_{12}-\alpha^{\ast}\sigma_{21}\right)  ,\nonumber\\
\partial_{t}\rho_{11}  &  =-\gamma_{10}\rho_{11}+\gamma_{21}\rho_{22}+i\left(
\alpha\sigma_{12}-\alpha^{\ast}\sigma_{21}\right)  ,\\
\partial_{t}\rho_{00}  &  =\gamma_{10}\rho_{11}+\gamma_{20}\rho_{22}%
-R\rho_{00}\\
\partial_{t}\sigma_{12}  &  =-\left(  \gamma_{\bot}+i\delta\right)
\sigma_{21}+i\alpha^{\ast}\left(  \rho_{22}-\rho_{11}\right)  .
\end{align}
Now we must take into account that the lower lasing level $\left\vert
1\right\rangle $ usually relaxes very fast towards level $\left\vert
0\right\rangle $. This means that $\rho_{11}\approx0$ and consequently that
$d=\rho_{22}-\rho_{11}\approx\rho_{22}$. Taking this into account and also
that $\rho_{00}+\rho_{22}\approx1$ in this approximation, we are left with
\begin{align}
\partial_{t}\rho_{22}  &  =-\left(  \gamma_{20}+\gamma_{21}+R\right)  d+R\\
&  +i\left(  \alpha\sigma_{12}-\alpha^{\ast}\sigma_{21}\right)  ,\nonumber\\
\partial_{t}\sigma_{12}  &  =-\left(  \gamma_{\bot}+i\delta\right)
\sigma_{21}+i\alpha^{\ast}d.
\end{align}

We see that, after the adiabatic elimination of $\rho_{33}$ and $\rho_{11}$,
the four--level Bloch equations are isomorphic to Eqs. (\ref{B1}) and
(\ref{B2}) that describe two--level atoms. Then we can apply the two--level
description to a four--level atom by making the following identifications%
\begin{align}
\gamma_{||}  &  \rightarrow\gamma_{20}+\gamma_{21}+R,\label{gammapar4L}\\
d_{0}  &  \rightarrow\frac{R}{\gamma_{20}+\gamma_{21}+R}. \label{pump4L}%
\end{align}
Again, as it was the case for three--level lasers, the population decay rate
$\gamma_{||}$ and the pumping rate $d_{0}$ of the two--level theory must be
reinterpreted when applied to four--level lasers.

Once we have shown that the two--level theory of Eqs. (\ref{B1}) and
(\ref{B2}) can be applied to three-- and four--level lasers by suitably
interpreting the parameters $\gamma_{||}$ and $d_{0}$, in the following we
shall always refer to the two--level model but the reader must keep in mind
that the transformations we have derived must be taken into account when
applying this theory to three-- and four--level lasers.

\section{The Maxwell--Bloch Equations}

\label{Maxwell-Bloch equations}Once the field equation (\ref{maxE}) and the
optical Bloch equations for matter dynamics Eqs. (\ref{B1}) and (\ref{B2})
have been derived, we only need to connect them in order to obtain a closed
set of equations for the analysis of amplification and laser dynamics.

Under the action of the light field each atom develops an electric dipole. As
the number of atoms contained in a small volume (small as compared with the
light wavelength) is always large, one can assume that at each spatial
position $\mathbf{r}$ there exists a polarization given by the
quantum-mechanical expectation value of the electric dipole moment operator
$\boldsymbol{\hat{\mu}}$. When using the density (population)\ matrix
formalism, this expectation value is computed as the trace%
\begin{equation}
\mathbf{P}\left(  \mathbf{r},t\right)  =\mathcal{N}\mathrm{Tr}\left(
\rho\left(  \mathbf{r},t\right)  \boldsymbol{\hat{\mu}}\right)  ,
\end{equation}
where $\mathcal{N}$ denotes the number of atoms per unit volume. Making use of
Eq. (\ref{rho}) and of the matrix form for the dipole moment operator%
\[
\boldsymbol{\hat{\mu}}=%
\begin{bmatrix}
0 & \boldsymbol{\mu}_{21}\\
\boldsymbol{\mu}_{12} & 0
\end{bmatrix}
,
\]
one has%
\begin{equation}
\mathbf{P}\left(  \mathbf{r},t\right)  =\mathcal{N}\left[  \boldsymbol{\mu
}_{12}\rho_{21}\left(  \mathbf{r},t\right)  +c.c.\right]  ,
\end{equation}
which, making use of definitions (\ref{s12-s21}), reads%
\begin{equation}
\mathbf{P}\left(  \mathbf{r},t\right)  =\mathcal{N}\left[  \boldsymbol{\mu
}_{12}\sigma_{21}\left(  \mathbf{r},t\right)  e^{i\left(  kz-\omega t\right)
}+c.c.\right]  ,
\end{equation}
which, compared with Eq. (\ref{P}) yields%
\begin{equation}
\mathcal{P}\left(  z,t\right)  =2\mathcal{N}\left(  \boldsymbol{\mu}_{12}%
\cdot\mathbf{e}^{\ast}\right)  \sigma_{21}\left(  \mathbf{r},t\right)  .
\label{Prho}%
\end{equation}
We finally come back to the wave equation (\ref{maxE}), multiply it by
$\left(  \boldsymbol{\mu}_{21}\cdot\mathbf{e}\right)  /2\hbar$, and make use
of Eqs. (\ref{alpha}) and (\ref{Prho}) for obtaining the final field equation,
which we write together with the Bloch equations (\ref{B1})--(\ref{B2}) for
the sake of convenience
\begin{gather}
\frac{\partial\alpha}{\partial t}+c\frac{\partial\alpha}{\partial z}%
=ig\sigma_{21},\label{WE}\\
\partial_{t}\sigma_{12}=-\left(  \gamma_{\bot}+i\delta\right)  \sigma
_{12}+i\alpha^{\ast}d,\label{MBloc1}\\
\partial_{t}d=\gamma_{||}\left(  d_{0}-d\right)  +2i\left(  \alpha\sigma
_{12}-\alpha^{\ast}\sigma_{21}\right)  , \label{MBloch2}%
\end{gather}
where we have introduced the radiation--matter coupling constant%
\begin{equation}
g=\frac{\mathcal{N}\omega\left\vert \boldsymbol{\mu}_{21}\cdot\mathbf{e}%
\right\vert ^{2}}{2\varepsilon_{0}\hbar}\text{.} \label{g}%
\end{equation}
Note that Eqs. (\ref{WE}--\ref{MBloch2}), form a closed set of equations that
completely determines, self-consistently, the interaction between a light
field (of amplitude proportional to $\alpha$, see Eq. (\ref{alpha})) and a
collection of two--level atoms. This set of equations is known as the
\textit{Maxwell--Bloch equations for a two--level system}, which can be
applied to three-- and four--level systems by introducing the parameter
changes (\ref{gammapar3L},\ref{pump3L}) and (\ref{gammapar4L},\ref{pump4L}), respectively.

\section{Amplification}

\label{amplification}The simplest issue that can be studied within the
developed formalism is the amplification of a monochromatic light beam after
traveling some distance along a medium. If we identify $\omega$ with the
actual light frequency, then $\mathcal{E}\left(  z,t\right)  =\mathcal{E}%
\left(  z\right)  $, see Eq. (\ref{E}), which implies that $\alpha\left(
z,t\right)  =\alpha\left(  z\right)  $. On the other hand, after a short
transient (of the order of the inverse of the decay constants), the atomic
system will have reached a steady configuration, which is ensured by the
presence of damping. Thus, after that transient one can ignore the time
derivatives in the Maxwell--Bloch equations. Solving for the material
variables (\ref{B1})--(\ref{B2}) in steady state, one has%
\begin{align}
d_{\mathrm{s}}  &  =d_{0}\frac{\gamma_{\bot}^{2}+\delta^{2}}{\delta^{2}%
+\gamma_{\bot}^{2}+4\gamma_{\bot}\left\vert \alpha\right\vert ^{2}/\gamma
_{||}},\label{ds}\\
\sigma_{21,\mathrm{s}}  &  =d_{0}\alpha\frac{\delta-i\gamma_{\bot}}{\delta
^{2}+\gamma_{\bot}^{2}+4\gamma_{\bot}\left\vert \alpha\right\vert ^{2}%
/\gamma_{||}}, \label{sigmas}%
\end{align}
where the subscript "$\mathrm{s}$" refers to the steady state. Substituting
the result into the field equation (\ref{WE}) one has%
\begin{equation}
\frac{d\alpha}{dz}=\frac{d_{0}g}{c}\frac{\gamma_{\bot}+i\delta}{\delta
^{2}+\gamma_{\bot}^{2}+4\gamma_{\bot}\left\vert \alpha\right\vert ^{2}%
/\gamma_{||}}\alpha. \label{dadz}%
\end{equation}
This equation governs the spatial variation of the field amplitude $\alpha$
along the atomic medium.

\subsection{Weak field limit}

Before considering the general solution let us concentrate first on the weak
field limit, defined as $\left\vert \alpha\right\vert ^{2}\ll\gamma_{\bot
}\gamma_{||}/4$. In this case the last term of the denominator in Eq.
(\ref{dadz}) can be ignored and the solution reads%
\begin{equation}
\alpha\left(  z\right)  =\alpha\left(  0\right)  \exp\left[  \frac
{a/2}{1+\left(  \delta/\gamma_{\bot}\right)  ^{2}}\ \left(  1+i\frac{\delta
}{\gamma_{\bot}}\right)  z\right]  , \label{amplin}%
\end{equation}
where%
\begin{equation}
a=\frac{2d_{0}g}{c\gamma_{\bot}}, \label{a}%
\end{equation}
and $g$ given by Eq. (\ref{g}). Parameter $a$ is responsible for the
attenuation (when $a<0$, i.e., when $d_{0}<0$) or amplification ($a>0$, i.e.,
$d_{0}>0$) of the light along its propagation through the material. In case of
attenuation, the inverse $a^{-1}$ is known as \textit{penetration depth}. In
case of amplification $a$ receives the name of \textit{small-signal gain per
unit length.} (Note that for $\delta=0$, $\left\vert \alpha\left(  z\right)
\right\vert ^{2}=\left\vert \alpha\left(  0\right)  \right\vert ^{2}%
\exp\left(  az\right)  $.)

On the other hand the imaginary exponent corresponds to a correction to the
light wavenumber. In fact, noticing that $\alpha$ is proportional to the field
amplitude $\mathcal{E}$ and recalling Eq. (\ref{E}) one has that the actual
wavenumber is%
\begin{equation}
k_{\mathrm{eff}}=k+\delta k,
\end{equation}
and consequently the refractive index $n=ck_{\mathrm{eff}}/\omega$ reads%
\begin{align}
n  &  =\frac{ck}{\omega}+\frac{d_{0}g}{\gamma_{\bot}\omega}\frac{\left(
\delta/\gamma_{\bot}\right)  }{1+\left(  \delta/\gamma_{\bot}\right)  ^{2}%
}\nonumber\\
&  =1+\frac{d_{0}\mathcal{N}\left\vert \boldsymbol{\mu}_{21}\cdot
\mathbf{e}\right\vert ^{2}}{2\varepsilon_{0}\hbar\gamma_{\bot}}\frac{\left(
\delta/\gamma_{\bot}\right)  }{1+\left(  \delta/\gamma_{\bot}\right)  ^{2}},
\end{align}
which has the same qualitative behavior as the classical expression obtained
from the (harmonic oscillator) Lorentz model \cite{MilonniEberly}.

\subsection{Strong field limit}

In the opposite limit, namely $\left\vert \alpha\right\vert ^{2}\gg
\gamma_{\bot}\gamma_{||}/4,\delta^{2}$, Eq. (\ref{dadz}) becomes%
\begin{equation}
\frac{d\alpha}{dz}=\frac{\gamma_{||}d_{0}g}{4c}\left(  1+i\frac{\delta}%
{\gamma_{\bot}}\right)  \frac{\alpha}{\left\vert \alpha\right\vert ^{2}}.
\end{equation}
Multiplying this equation by $\alpha^{\ast}$ and taking the real part of the
resulting equation one has%
\begin{equation}
\frac{d\left\vert \alpha\right\vert ^{2}}{dz}=\frac{\gamma_{||}d_{0}g}%
{2c}=\frac{\gamma_{||}\gamma_{\bot}a}{4},
\end{equation}
whose solution reads%
\begin{equation}
\left\vert \alpha\left(  z\right)  \right\vert ^{2}=\left\vert \alpha\left(
0\right)  \right\vert ^{2}+\frac{\gamma_{||}\gamma_{\bot}a}{4}z.
\label{strong}%
\end{equation}
Again, amplification requires $a>0$, i.e., $d_{0}>0$. This result means that,
for strong fields, there exists saturation: the amplification (whenever
$d_{0}>0$) persists but it is \textit{linear} in the propagation distance,
differently from the weak field limit, in which amplification occurs
exponentially, see Eq. (\ref{amplin}).

\subsection{General solution}

In order to consider the general case it is convenient to use a polar
decomposition for $\alpha$ as%
\begin{equation}
\alpha\left(  z\right)  =\left\vert \alpha\left(  z\right)  \right\vert
e^{i\phi\left(  z\right)  }. \label{polar}%
\end{equation}
Substituting this expression into Eq. (\ref{dadz}) and separating it into its
real and imaginary parts one obtains%
\begin{align}
\frac{d\left\vert \alpha\right\vert }{dz}  &  =\frac{\gamma_{\bot}d_{0}g}%
{c}\frac{\left\vert \alpha\right\vert }{\delta^{2}+\gamma_{\bot}^{2}%
+4\gamma_{\bot}^{2}\left\vert \alpha\right\vert ^{2}/\gamma_{||}}%
,\label{dAdz}\\
\frac{d\phi}{dz}  &  =\frac{\delta d_{0}g}{c}\frac{1}{\delta^{2}+\gamma_{\bot
}^{2}+4\gamma_{\bot}^{2}\left\vert \alpha\right\vert ^{2}/\gamma_{||}}.
\label{dfidz}%
\end{align}
\begin{figure}[t]
\begin{center}
\scalebox{0.8}{\includegraphics{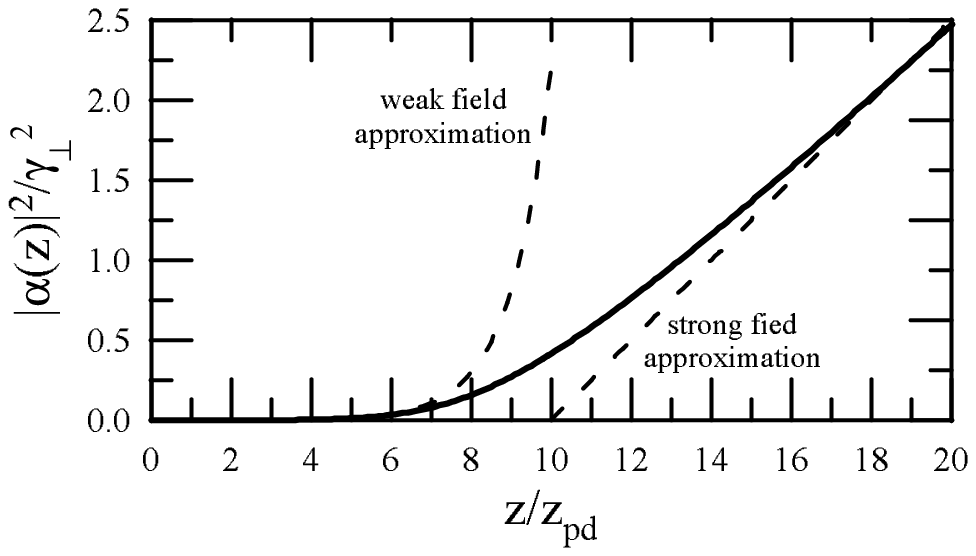}}
\end{center}
\caption{Field intensity during amplification as a function of the
normalized distance $z/z_{pd}$, with $z_{pd}=a^{-1}$ the penetration
depth. The dashed lines correspond to the weak and strong field
approximations, Eqs. (\ref{amplin}) and (\ref{strong}),
respectively.}
\end{figure}

Eq. (\ref{dAdz}) can be integrated to yield%
\begin{equation}
\left(  \gamma_{\bot}^{2}+\delta^{2}\right)  \ln\frac{\left\vert \alpha\left(
z\right)  \right\vert }{\left\vert \alpha\left(  0\right)  \right\vert
}+2\frac{\gamma_{\bot}}{\gamma_{||}}\left[  \left\vert \alpha\left(  z\right)
\right\vert ^{2}-\left\vert \alpha\left(  0\right)  \right\vert ^{2}\right]
=\frac{\gamma_{\bot}d_{0}g}{c}z, \label{eqA}%
\end{equation}
which does not allow an explicit expression for $\left\vert \alpha\left(
z\right)  \right\vert $. In any case, Eq. (\ref{dAdz}) shows that $\left\vert
\alpha\left(  z\right)  \right\vert ^{-1}d\left\vert \alpha\left(  z\right)
\right\vert /dz$ has the same sign as $d_{0}$, so that $d_{0}>0$ implies
amplification. In Fig. 6 the solution of Eq. (\ref{eqA}) is represented as a
function of $z$ together with the weak and strong field approximations derived above.

As for Eq. (\ref{dfidz}), the phase can be determined by noticing that%
\begin{equation}
\frac{d\phi}{d\left\vert \alpha\right\vert }=\frac{d\phi/dz}{d\left\vert
\alpha\right\vert /dz}=\frac{\delta}{\gamma_{\bot}}\frac{1}{\left\vert
\alpha\right\vert }, \label{dphi}%
\end{equation}
from which%
\begin{equation}
\phi=\phi_{0}+\frac{\delta}{\gamma_{\bot}}\ln\frac{\left\vert \alpha\left(
z\right)  \right\vert }{\left\vert \alpha\left(  0\right)  \right\vert }.
\label{fase}%
\end{equation}
Note that, on resonance ($\delta=\omega-\omega_{21}=0$) there is no phase
variation along the propagation direction (apart from the original phase $kz$).

\section{Lasing}

Differently from the previous analysis, in which we assumed that a given field
(whose frequency and initial amplitude are known data) is injected into the
entrance face of a material, the light field in a laser is not fixed
externally but is self-consistently generated by the medium, through
amplification, and must verify the boundary conditions imposed by the cavity.
As the model we have developed considers a traveling wave (moving in one
direction) the following analysis only applies to ring lasers in which
unidirectional operation can take place (in linear, i.e. Fabry--Perot,
resonators there are two counterpropagating waves that form a standing wave, a
more complicated case that we shall not treat here).

\subsection{Boundary condition}
\begin{figure}[t]
\begin{center}
\scalebox{0.8}{\includegraphics{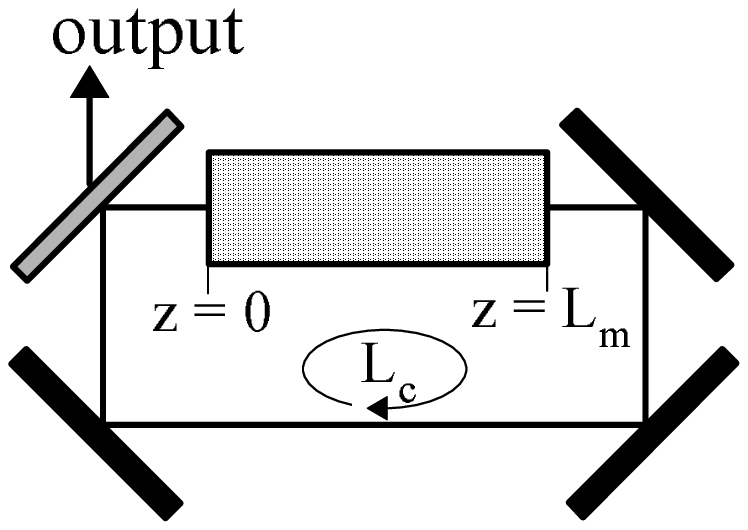}}
\end{center}
\caption{Scheme of the ring cavity. The active medium is placed in
the region $0<z<L_{\mathrm{m}}$. The black cavity mirrors are
perfectly reflecting whilst the grey mirror has a finite
reflectivity $\mathcal{R}$. The arrows indicate the propagation of
the intracavity and output fields.}
\end{figure}
We assume that the medium is of length $L_{\mathrm{m}}$ and that the cavity
has a length $L_{\mathrm{c}}$, see Fig. 7. We take $z=0$ as the entrance face
of the amplifying medium. The boundary condition imposed by the resonator
reads%
\begin{equation}
\mathbf{E}\left(  0,t\right)  =\mathcal{R}\mathbf{E}\left(  L_{\mathrm{m}%
},t-\Delta t\right)  , \label{BC1}%
\end{equation}
where $\mathcal{R}$ represents the (amplitude) reflectivity of the mirrors
($0\leq\mathcal{R}^{2}\leq1$ gives the fraction of light power that survives
after a complete cavity round trip) and%
\begin{equation}
\Delta t=\frac{L_{\mathrm{c}}-L_{\mathrm{m}}}{c} \label{Dt}%
\end{equation}
is the time delay taken by the light to travel from the exit face of the
medium back to its entrance face after being reflected by the cavity mirrors.

Making use of Eq. (\ref{E}), and after little algebra, the boundary condition
(\ref{BC1}) reads%
\begin{equation}
\mathcal{E}\left(  0,t\right)  =\mathcal{R}\exp\left[  i\left(  kL_{\mathrm{m}%
}+\omega\Delta t\right)  \right]  \mathcal{E}\left(  L_{\mathrm{m}},t-\Delta
t\right)  ,
\end{equation}
which, upon using Eq. (\ref{Dt}) and recalling that $k=\omega/c$ (this was our
choice in writing Eq. (\ref{E})), reads%
\begin{equation}
\mathcal{E}\left(  0,t\right)  =\mathcal{R}e^{ikL_{\mathrm{c}}}\mathcal{E}%
\left(  L_{\mathrm{m}},t-\Delta t\right)  .
\end{equation}
Finally, multiplying this equation by $\left(  \boldsymbol{\mu}_{21}%
\cdot\mathbf{e}\right)  /2\hbar$ and recalling Eq. (\ref{alpha}), one has%
\begin{equation}
\alpha\left(  0,t\right)  =\mathcal{R}e^{ikL_{\mathrm{c}}}\alpha\left(
L_{\mathrm{m}},t-\Delta t\right)  . \label{BC2}%
\end{equation}
We analyze next the monochromatic lasing solution.

\subsection{Monochromatic (singlemode) emission}

We note that the frequency $\omega$ appearing in the field expression
(\ref{E}) is by now unknown. Under monochromatic operation the laser light
has, by definition, a single frequency. \textit{If we take} $\omega$
\textit{to be the actual lasing mode frequency}, the field amplitude must be
then a constant in time i.e., $\alpha\left(  z,t\right)  =\alpha\left(
z\right)  $, as in the previous analysis. Thus Eq. (\ref{BC2}) becomes%
\begin{equation}
\alpha\left(  0\right)  =\mathcal{R}e^{ikL_{\mathrm{c}}}\alpha\left(
L_{\mathrm{m}}\right)  .
\end{equation}
Using now the polar decomposition (\ref{polar}) one has%
\begin{align}
\left\vert \alpha\left(  0\right)  \right\vert ^{2}  &  =\mathcal{R}%
^{2}\left\vert \alpha\left(  L_{\mathrm{m}}\right)  \right\vert ^{2}%
,\label{BCA}\\
\phi\left(  0\right)   &  =\phi\left(  L_{\mathrm{m}}\right)  +kL_{\mathrm{c}%
}+2m\pi, \label{BCphi}%
\end{align}
being $m$ an integer.

\subsubsection{Determination of the laser intensity}

Let us first analyze the laser intensity $\left\vert \alpha\right\vert ^{2}$.
(In fact the laser intensity is proportional to $\left\vert \mathcal{E}%
\right\vert ^{2}$ but remind that $\mathcal{E}\propto\alpha$, and then
$\left\vert \mathcal{E}\right\vert ^{2}\propto\left\vert \alpha\right\vert
^{2}$.) We note that, as we are dealing with a field whose amplitude is time
independent, the analysis of amplification performed in the previous section
is directly applicable. Making use of Eq. (\ref{BCA}), Eq. (\ref{eqA})
becomes, for $z=L_{\mathrm{m}}$,%
\begin{align}
\frac{\gamma_{\bot}d_{0}g}{c}L_{\mathrm{m}}  &  =\frac{1}{2}\left(
\gamma_{\bot}^{2}+\delta^{2}\right)  \ln\mathcal{R}^{-2}\nonumber\\
&  +2\frac{\gamma_{\bot}}{\gamma_{||}}\left(  1-\mathcal{R}^{2}\right)
\left\vert \alpha\left(  L_{\mathrm{m}}\right)  \right\vert ^{2},
\end{align}
which, after trivial manipulation yields%
\begin{equation}
\left\vert \alpha\left(  L_{\mathrm{m}}\right)  \right\vert ^{2}=\frac
{\gamma_{||}\gamma_{\bot}}{4}\frac{\left\vert \ln\mathcal{R}^{2}\right\vert
}{1-\mathcal{R}^{2}}\left(  r-1-\Delta^{2}\right)  , \label{stat}%
\end{equation}
where we made $\ln\mathcal{R}^{-2}=\left\vert \ln\mathcal{R}^{2}\right\vert $
(remind that $\mathcal{R}^{2}\leq1$) and we have defined two important
parameters, the adimensional pump $r$ and the normalized detuning $\Delta$
through%
\begin{align}
r  &  =\frac{2d_{0}gL_{\mathrm{m}}}{\gamma_{\bot}c\left\vert \ln
\mathcal{R}^{2}\right\vert }=\frac{aL_{\mathrm{m}}}{\left\vert \ln
\mathcal{R}^{2}\right\vert },\label{r}\\
\Delta &  =\frac{\delta}{\gamma_{\bot}}=\frac{\omega-\omega_{21}}{\gamma
_{\bot}}. \label{Delta}%
\end{align}
We note that the adimensional parameter $r$ is proportional to the gain
properties of the medium and inversely proportional to the damping properties
of the system. In fact, $aL_{\mathrm{m}}$ gives the small-signal single-pass
gain along the amplifying medium (remind that $a$, Eq. (\ref{a}), is the
small-signal gain per unit length). Thus $r$ acts as an effective pumping
parameter, as will become clear next. Equation (\ref{stat}) determines the
value of the field intensity at the exit face of the amplifying medium.
Clearly, in order to be meaningful, $\left\vert \alpha\left(  L_{\mathrm{m}%
}\right)  \right\vert ^{2}\geq0$, what implies%
\begin{equation}
r\geq r_{on}\equiv1+\Delta^{2}. \label{threshold}%
\end{equation}
Thus parameter $r$ must exceed a given threshold (the lasing threshold $r_{0}%
$) in order that the laser emits light. This is why $r$ is called the pump
parameter (there is a minimum pump required for the system starts lasing).
\begin{figure}[t]
\begin{center}
\scalebox{0.8}{\includegraphics{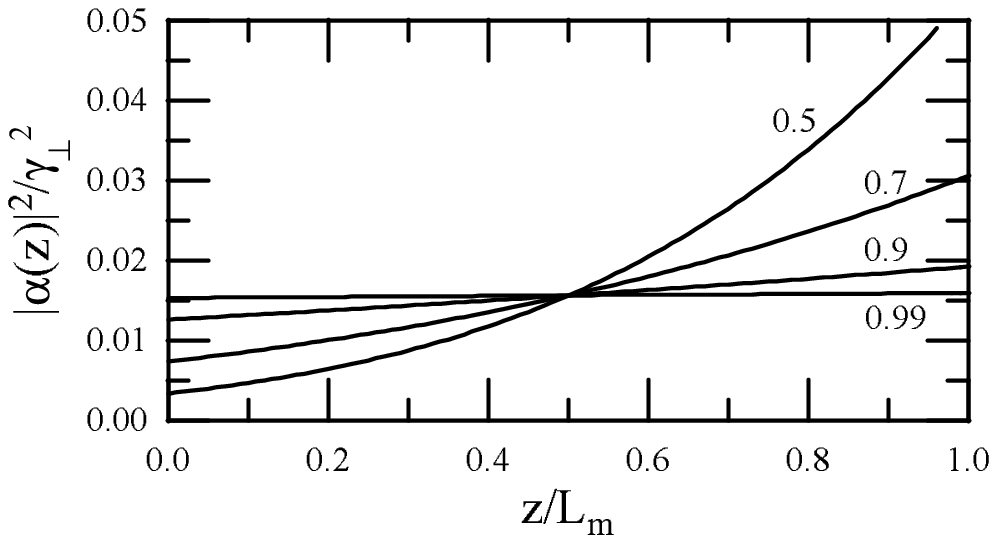}}
\end{center}
\caption{Intracavity field intensity as a function of distance for
$\gamma_{||}=\gamma_{\bot},\Delta=0,r=1.5$ and the values of
$\mathcal{R}$ marked in the figure.}
\end{figure}
What we have obtained is the field intensity at the faces of the active
medium, Eqs. (\ref{BCA}) and (\ref{stat}). But it is also interesting to
analyze how this intensity varies along the active medium. Thus, after using
Eqs. (\ref{BCA}), (\ref{r}) and (\ref{Delta}), we write down Eq. (\ref{eqA})
in the form%
\begin{gather}
r\left\vert \ln\mathcal{R}^{2}\right\vert \frac{z}{L_{\mathrm{m}}}=\left(
1+\Delta^{2}\right)  \ln\frac{\left\vert \alpha\left(  z\right)  \right\vert
^{2}}{\mathcal{R}^{2}\left\vert \alpha\left(  L_{\mathrm{m}}\right)
\right\vert ^{2}}\nonumber\\
+\frac{4}{\gamma_{||}\gamma_{\bot}}\left[  \left\vert \alpha\left(  z\right)
\right\vert ^{2}-\mathcal{R}^{2}\left\vert \alpha\left(  L_{\mathrm{m}%
}\right)  \right\vert ^{2}\right]  ,
\end{gather}
with $\left\vert \alpha\left(  L_{\mathrm{m}}\right)  \right\vert ^{2}$ given
by Eq. (\ref{stat}). This equation can be solved numerically and in Fig. 8 we
represent its solutions for fixed parameters and several values of the
reflectivity $\mathcal{R}^{2}$, showing that as $\mathcal{R}^{2}$ approaches
unity the solution becomes progressively uniform. This fact suggests that for
$\mathcal{R}^{2}\rightarrow1$, it must be possible to rewrite the laser
equations in a simpler way as in this limit the steady state is independent of
$z$. We shall come back to this point in the next section. But first we shall
continue analyzing the laser steady state.

\subsubsection{Determination of the laser frequency}

Even if it can seem that we know the lasing intensity value, the fact is that
we still ignore the value of the lasing frequency $\omega$ and thus the value
of $\Delta$. This problem is solved by considering the phase boundary
condition (\ref{BCphi}). First we recall Eq. (\ref{fase}), which we write in
the form%
\begin{equation}
\phi\left(  L_{\mathrm{m}}\right)  -\phi\left(  0\right)  =\tfrac{1}{2}%
\Delta\ln\frac{\left\vert \alpha\left(  L_{\mathrm{m}}\right)  \right\vert
^{2}}{\left\vert \alpha\left(  0\right)  \right\vert ^{2}}=-\tfrac{1}{2}%
\Delta\left\vert \ln\mathcal{R}^{2}\right\vert , \label{BCphi2}%
\end{equation}
where Eq. (\ref{BCA}) has been used in the last equality. Comparison between
Eqs. (\ref{BCphi}) and (\ref{BCphi2}) yields%
\begin{equation}
\tfrac{1}{2}\Delta\left\vert \ln\mathcal{R}^{2}\right\vert =2\pi
m-kL_{\mathrm{c}}. \label{eqD1}%
\end{equation}
We now introduce the wavenumber $k_{\mathrm{c}}$ and frequency $\omega
_{\mathrm{c}}$ of the cavity longitudinal mode closest to the atomic
resonance. As we are dealing with a cavity longitudinal mode, it must be
verified, by definition, that%
\begin{equation}
k_{\mathrm{c}}=2\pi m_{\mathrm{c}}/L_{\mathrm{c}},\qquad\omega_{\mathrm{c}%
}=ck_{\mathrm{c}}, \label{cavitymode}%
\end{equation}
being $m_{\mathrm{c}}$ an integer. Substituting these quantities into Eq.
(\ref{eqD1}) one gets%
\begin{equation}
\tfrac{1}{2}\Delta\left\vert \ln\mathcal{R}^{2}\right\vert =\left(
\omega_{\mathrm{c}}-\omega\right)  \frac{L_{\mathrm{c}}}{c}+2\pi n,
\label{eqD2}%
\end{equation}
where $n=m-m_{\mathrm{c}}$ is a new integer. We finally recall Eq.
(\ref{Delta}) so that Eq. (\ref{eqD2}) yields the following value for the
laser frequency%
\begin{equation}
\omega_{n}=\frac{\kappa\omega_{21}+\gamma_{\bot}\omega_{\mathrm{c}}}%
{\kappa+\gamma_{\bot}}+n\frac{\gamma_{\bot}}{\kappa+\gamma_{\bot}}\frac{2\pi
c}{L_{\mathrm{c}}} \label{pulling}%
\end{equation}
where we have defined%
\begin{equation}
\kappa=\frac{c\left\vert \ln\mathcal{R}^{2}\right\vert }{2L_{\mathrm{c}}},
\label{kappa}%
\end{equation}
which is known as the \textit{cavity damping rate} for reasons that will be
analyzed in the next section. We note that Eq. (\ref{pulling}) indicates that
there exists a family of solutions (labeled by the integer $n$). As we show
next all these solutions have, in general, different lasing thresholds. From
Eq. (\ref{pulling}) the lasing threshold (\ref{threshold}) can be finally
determined as%
\begin{equation}
r_{on}=1+\left(  \frac{\omega_{\mathrm{c}}-\omega_{21}+n\frac{2\pi
c}{L_{\mathrm{c}}}}{\kappa+\gamma_{\bot}}\right)  ^{2}.
\end{equation}
Now, the difference between the cavity and atomic transition frequencies is
obviously smaller than the free spectral range, i.e., $\left\vert
\omega_{\mathrm{c}}-\omega_{21}\right\vert <2\pi c/L_{\mathrm{c}}$. This makes
that $r_{on}$ is minimum for $n=0$ and, also, that the frequency of the
amplified mode, $\omega_{0}$, be given by
\begin{equation}
\omega_{0}=\frac{\kappa\omega_{21}+\gamma_{\bot}\omega_{\mathrm{c}}}%
{\kappa+\gamma_{\bot}},
\end{equation}
which is the \textit{pulling formula}. The result is that the laser frequency
is a compromise between the cavity and atomic transition frequency. Notice
that for a "good cavity", $\kappa\ll\gamma_{\bot}$, the laser frequency
approaches the cavity frequency, whilst for a "bad cavity", $\gamma_{\bot}%
\ll\kappa$, the laser frequency approaches that of the atomic transition. This
is a quite intuitive result indeed.

\subsubsection{The resonant case}

\label{resonant}Let us analyze the relevant case $\omega_{\mathrm{c}}%
=\omega_{21}$, corresponding to a cavity exactly tuned to the atomic
resonance. In this case the pump must verify%
\begin{equation}
r\geq r_{on}=1+\left[  \frac{2\pi c}{\left(  \kappa+\gamma_{\bot}\right)
L_{\mathrm{c}}}\right]  ^{2}n^{2},
\end{equation}
and the lasing mode with lowest threshold is that with $n=0$, as discussed
above. Hence, \textit{at resonance}, the basic lasing solution has a threshold
given by $r_{on}=1$, and its frequency is $\omega=\omega_{\mathrm{c}}%
=\omega_{21}$, see Eq. (\ref{pulling}) for $n=0$.

The amplitude of this lasing solution verifies Eq. (\ref{dadz}) with
$\delta=\omega-\omega_{21}=0$:%
\begin{equation}
\frac{d\alpha_{\mathrm{s}}}{dz}=\frac{d_{0}g}{\gamma_{\bot}c}\frac{1}%
{1+\frac{4}{\gamma_{\bot}\gamma_{||}}\left\vert \alpha_{\mathrm{s}}\right\vert
^{2}}\alpha_{\mathrm{s}}. \label{dadzres}%
\end{equation}
We note that we have introduced the subscript "$\mathrm{s}$" to emphasize that
this amplitude $\alpha$ corresponds to the steady lasing solution.

Finally, the "intensity" of the laser light at the exit of the active medium
is given by Eq. (\ref{stat}) with $\Delta=\delta/\gamma_{\bot}=0$:%
\begin{equation}
\left\vert \alpha_{\mathrm{s}}\left(  L_{\mathrm{m}}\right)  \right\vert
^{2}=\frac{\gamma_{||}\gamma_{\bot}}{4}\frac{\left\vert \ln\mathcal{R}%
^{2}\right\vert }{1-\mathcal{R}^{2}}\left(  r-1\right)  . \label{statres}%
\end{equation}
We note that there exists no phase variation of the laser complex amplitude
$\alpha_{\mathrm{s}}$ along the medium, see Eq. (\ref{dfidz}) with $\delta=0$.
We shall make use of these expressions in the following section.

\section{The laser equations in the uniform field limit}

In this section we want to find a simpler model that allows us to study laser
dynamics and instabilities in an easy way. The sought model is known as the
Lorenz--Haken model and can be rigorously derived from the Maxwell--Bloch
equations (\ref{B1})--(\ref{B2}) and (\ref{WE}) in the so-called uniform field
limit, which we consider now. This limit assumes that the cavity reflectivity
is closest to unity ($\mathcal{R}^{2}\rightarrow1$ in all previous
expressions). For the sake of simplicity \cite{nonresonant} the derivation
will be done in the resonant case, where the cavity is tuned in such a way
that one of its longitudinal modes has a frequency $\omega_{\mathrm{c}}$ that
matches exactly the atomic resonance frequency $\omega_{21}$. In this case the
analysis done in Sec. (\ref{resonant}) suggests to choose the value of the
arbitrary frequency $\omega$ as $\omega=\omega_{\mathrm{c}}=\omega_{21}$. (We
remind that we can choose freely this value. If this election is "wrong" the
laser equations will yield an electric field amplitude $\alpha$ which contains
a phase factor of the form $\exp\left(  -i\Delta\omega t\right)  $ that will
define the actual laser frequency.)

First we recall the Maxwell--Bloch equations (\ref{WE}--\ref{MBloch2}) for
$\delta=0$:%
\begin{gather}
\frac{\partial\alpha}{\partial t}+c\frac{\partial\alpha}{\partial z}%
=ig\sigma_{21},\label{MB1}\\
\frac{\partial\sigma_{21}}{\partial t}=-\gamma_{\bot}\sigma_{21}-i\alpha
d,\label{MB2}\\
\frac{\partial d}{\partial t}=\gamma_{||}\left(  d_{0}-d\right)  +2i\left(
\alpha\sigma_{12}-\alpha^{\ast}\sigma_{21}\right)  , \label{MB3}%
\end{gather}
which are to be supplemented by the boundary condition (\ref{BC2}) with
$k=k_{\mathrm{c}}=\omega_{\mathrm{c}}/c$, see Eq. (\ref{dispersion0}), so that
$k=2\pi m_{\mathrm{c}}/L_{\mathrm{c}}$, see Eq. (\ref{cavitymode}). With these
assumptions the boundary condition (\ref{BC2}) becomes%
\begin{equation}
\alpha\left(  0,t\right)  =\mathcal{R}\alpha\left(  L_{\mathrm{m}},t-\Delta
t\right)  . \label{BCMB}%
\end{equation}
We note that this boundary condition is not isochronous (it relates values of
the field amplitude at different times) and this makes difficult the analysis.
We note for later use that this boundary condition applies, in particular, to
the steady lasing solution (independent of time) so that%
\begin{equation}
\alpha_{\mathrm{s}}\left(  0\right)  =\mathcal{R}\alpha_{\mathrm{s}}\left(
L_{\mathrm{m}}\right)  . \label{BCs}%
\end{equation}
These equations form the basis of our study.

\subsection{A first change of variables}

In order to make the boundary condition isochronous, we introduce the
following change of variables \cite{NarducciAbraham}:%
\begin{align}
\alpha^{\prime}\left(  z,t\right)   &  =\alpha\left(  z,t-\tau\right)  ,\ \ \\
\sigma_{21}^{\prime}\left(  z,t\right)   &  =\sigma_{21}\left(  z,t-\tau
\right)  ,\ \ \\
d^{\prime}\left(  z,t\right)   &  =d\left(  z,t-\tau\right)  ,
\end{align}
with $\tau=z\Delta t/L_{\mathrm{m}}$ and $\Delta t=\left(  L_{\mathrm{c}%
}-L_{\mathrm{m}}\right)  /c$, Eq. (\ref{Dt}). The new variables verify%
\begin{align}
\frac{\partial\alpha}{\partial t}  &  =\frac{\partial\alpha^{\prime}}{\partial
t},\label{cambio1}\\
\frac{\partial\alpha}{\partial z}  &  =\frac{\partial\alpha^{\prime}}{\partial
z}+\frac{\Delta t}{L_{\mathrm{m}}}\frac{\partial\alpha^{\prime}}{\partial t},
\label{cambio2}%
\end{align}
and similar expressions for the material variables. Substitution of the
previous relations into Eqs. (\ref{MB1})--(\ref{MB3}) yields%
\begin{gather}
\frac{L_{\mathrm{c}}}{L_{\mathrm{m}}}\frac{\partial\alpha^{\prime}}{\partial
t}+c\frac{\partial\alpha^{\prime}}{\partial z}=ig\sigma_{21}^{\prime
},\label{MB1prime}\\
\frac{\partial\sigma_{21}^{\prime}}{\partial t}=-\gamma_{\bot}\sigma
_{21}^{\prime}-i\alpha^{\prime}d^{\prime},\label{MB2prime}\\
\frac{\partial d^{\prime}}{\partial t}=-\gamma_{||}\left(  d^{\prime}%
-d_{0}\right)  +2i\alpha^{\prime}\sigma_{12}^{\prime}+c.c., \label{MB3prime}%
\end{gather}
where we used Eq. (\ref{Dt}). According to Eq. (\ref{BCMB}) the new variables
verify the following boundary condition%
\begin{equation}
\alpha^{\prime}\left(  0,t\right)  =\mathcal{R}\alpha^{\prime}\left(
L_{\mathrm{m}},t\right)  , \label{BCprime}%
\end{equation}
which is now \textit{isochronous}. We note that the definition of the new
variables is mathematically equivalent to "bend" the active medium on itself
so that its entrance ($z=0$) and exit faces ($z=L_{\mathrm{m}}$) coincide.

\subsection{A second change of variables}

Now we define another set of variables by referring the previous ones to their
monochromatic lasing values analyzed in the previous sections. The steady
values of the material variables have been calculated in Sec.
(\ref{amplification}), Eqs. (\ref{ds}) and (\ref{sigmas}), that,
particularized to the case $\delta=0$ we are considering, read%
\begin{align}
d_{\mathrm{s}}\left(  z\right)   &  =d_{0}\frac{1}{1+\frac{4}{\gamma_{\bot
}\gamma_{||}}\left\vert \alpha_{\mathrm{s}}\right\vert ^{2}},\label{dsres}\\
\sigma_{21,\mathrm{s}}\left(  z\right)   &  =\frac{-id_{0}\alpha_{\mathrm{s}}%
}{\gamma_{\bot}}\frac{1}{1+\frac{4}{\gamma_{\bot}\gamma_{||}}\left\vert
\alpha_{\mathrm{s}}\right\vert ^{2}}. \label{sigmasres}%
\end{align}
We note that these quantities are $z-$dependent as $\alpha_{\mathrm{s}}$ is,
Eq. (\ref{dadzres}). In particular we define the new variables through%
\begin{align}
F\left(  z,t\right)   &  =\sqrt{r-1}\frac{\alpha^{\prime}\left(  z,t\right)
}{\alpha_{\mathrm{s}}\left(  z\right)  },\;\label{FPD}\\
P\left(  z,t\right)   &  =\sqrt{r-1}\frac{\sigma_{21}^{\prime}\left(
z,t\right)  }{\sigma_{21,\mathrm{s}}\left(  z\right)  },\\
D\left(  z,t\right)   &  =\frac{d^{\prime}\left(  z,t\right)  }{d_{\mathrm{s}%
}\left(  z\right)  },
\end{align}
where $\alpha_{\mathrm{s}}$ verifies Eq. (\ref{dadzres}). The equations for
$F,P,D$ are obtained from Eqs. (\ref{MB1prime})--(\ref{MB3prime}). First the
equation for $F$ is computed. From its definition we have%
\begin{align}
\frac{\partial F}{\partial t}  &  =\sqrt{r-1}\frac{1}{\alpha_{\mathrm{s}}%
}\frac{\partial\alpha^{\prime}}{\partial t},\\
\frac{\partial F}{\partial z}  &  =\sqrt{r-1}\left(  \frac{1}{\alpha
_{\mathrm{s}}}\frac{\partial\alpha^{\prime}}{\partial z}-\frac{\alpha^{\prime
}}{\alpha_{\mathrm{s}}}\frac{1}{\alpha_{\mathrm{s}}}\frac{d\alpha_{\mathrm{s}%
}}{dz}\right)  .
\end{align}
Making use of these and of Eq. (\ref{dadzres}) we build the following equation
for $F$%
\begin{align}
\frac{L_{\mathrm{c}}}{L_{\mathrm{m}}}\frac{\partial F}{\partial t}%
+c\frac{\partial F}{\partial z}  &  =ig\frac{\sqrt{r-1}}{\alpha_{\mathrm{s}}%
}\sigma_{21}^{\prime}\nonumber\\
&  -\frac{d_{0}g}{\gamma_{\bot}}\frac{1}{1+\frac{4}{\gamma_{\bot}\gamma_{||}%
}\left\vert \alpha_{\mathrm{s}}\right\vert ^{2}}F.
\end{align}
that by using the definition of $P$ and of Eq. (\ref{sigmasres}) transforms
into%
\begin{equation}
\frac{\partial F}{\partial t}+v\frac{\partial F}{\partial z}=C_{F}\left(
z\right)  \left(  P-F\right)  , \label{dFdt}%
\end{equation}
where%
\begin{align}
v  &  =\frac{cL_{\mathrm{m}}}{L_{\mathrm{c}}},\\
C_{F}\left(  z\right)   &  =\frac{d_{0}g}{\gamma_{\bot}}\frac{L_{\mathrm{m}}%
}{L_{\mathrm{c}}}\frac{1}{1+\frac{4}{\gamma_{\bot}\gamma_{||}}\left\vert
\alpha_{\mathrm{s}}\right\vert ^{2}}. \label{CF}%
\end{align}
(Note that $v<c$ has the dimensions of a velocity.) The equations for the
material variables are easier to be obtained. Making use of the definitions of
$F$, $P$ and $D$, and making use of Eqs. (\ref{MB2prime}) and (\ref{MB3prime})
we obtain%
\begin{align}
\frac{\partial P}{\partial t}  &  =-\gamma_{\bot}P+C_{P}FD,\label{dPdt}\\
\frac{\partial D}{\partial t}  &  =-\gamma_{||}\left(  D-D_{0}\right)
-\left(  C_{D}FP^{\ast}+C_{D}^{\ast}F^{\ast}P\right)  , \label{dDdt}%
\end{align}
where%
\begin{align}
C_{P}\left(  z\right)   &  =-i\frac{\alpha_{\mathrm{s}}d_{\mathrm{s}}}%
{\sigma_{21,\mathrm{s}}},\;\\
D_{0}\left(  z\right)   &  =\frac{d_{0}}{d_{\mathrm{s}}},\;\\
C_{D}\left(  z\right)   &  =\frac{-2i\alpha_{\mathrm{s}}\sigma_{21,\mathrm{s}%
}^{\ast}}{d_{\mathrm{s}}\left(  r-1\right)  }.
\end{align}
After using the steady state equations (\ref{dsres}) and (\ref{sigmasres})
these expressions can be written as%
\begin{align}
C_{P}  &  =\gamma_{\bot},\label{CP}\\
D_{0}\left(  z\right)   &  =1+\frac{4}{\gamma_{\bot}\gamma_{||}}\left\vert
\alpha_{\mathrm{s}}\right\vert ^{2},\;C_{D}\left(  z\right)  =\frac
{2\left\vert \alpha_{\mathrm{s}}\right\vert ^{2}}{\gamma_{\bot}\left(
r-1\right)  }. \label{Cmat2}%
\end{align}
Up to this point, the equations for $F$, $P$, and $D$ are equivalent to the
original Maxwell--Bloch equations, as no approximation has been done.

\subsection{The Uniform Field Limit}

We now study the behavior of $C_{F}$, $C_{D}$ and $D_{0}$ in the case when the
cavity mirrors have a very good quality, i.e., when the reflectivity
$\mathcal{R}$ is very close to unity. In this limit, the boundary condition
(\ref{BCs}) says that $\alpha_{\mathrm{s}}\left(  0\right)  \approx
\alpha_{\mathrm{s}}\left(  L_{\mathrm{m}}\right)  $. On the other hand, the
steady state equation (\ref{dadzres}) tells us that $\left\vert \alpha
_{\mathrm{s}}\right\vert ^{2}$ is a monotonic growing function of $z$. Under
these circumstances, one can assume, to a very good approximation, that
$\left\vert \alpha_{\mathrm{s}}\left(  z\right)  \right\vert ^{2}$ is a
constant along the amplifying medium. (These facts can in fact be seen in Fig.
8.) In this case its value coincides, for instance, with its value at the
medium exit face, $\left\vert \alpha_{\mathrm{s}}\left(  L_{\mathrm{m}%
}\right)  \right\vert ^{2}$, which is given by Eq. (\ref{statres}):%
\[
\left\vert \alpha_{\mathrm{s}}\left(  z\right)  \right\vert ^{2}\approx
\frac{\gamma_{||}\gamma_{\bot}}{4}\frac{\left\vert \ln\mathcal{R}%
^{2}\right\vert }{1-\mathcal{R}^{2}}\left(  r-1\right)  ,\qquad\forall
z\text{.}%
\]
But, as we are considering the limit $\mathcal{R}\rightarrow1$, the quotient
$\left\vert \ln\mathcal{R}^{2}\right\vert /\left(  1-\mathcal{R}^{2}\right)  $
also tends to unity as can be checked easily, and we finally have%
\begin{equation}
\left\vert \alpha_{\mathrm{s}}\left(  z\right)  \right\vert ^{2}\approx
\frac{\gamma_{||}\gamma_{\bot}}{4}\left(  r-1\right)  .\qquad\forall z.
\label{intUFL}%
\end{equation}
This space uniformity of the laser intensity along the amplifying medium when
$\mathcal{R}\rightarrow1$ is the reason for the name "\textit{Uniform Field
Limit}". (We note that in the literature the uniform field limit has been
customarily associated not only with the high reflectivity condition but also
with the small gain condition $aL_{\mathrm{m}}\rightarrow0$. We see here that
the latter condition is completely superfluous.) Substitution of
(\ref{intUFL}) into Eqs. (\ref{CF}) and (\ref{Cmat2}) yields:%
\begin{align}
C_{F}\left(  z\right)   &  =\frac{d_{0}gL_{\mathrm{m}}}{\gamma_{\bot
}rL_{\mathrm{c}}},\nonumber\\
D_{0}\left(  z\right)   &  =r,\;C_{D}\left(  z\right)  =\frac{\gamma_{||}}%
{2},\qquad\forall z. \label{CDdef}%
\end{align}
Finally, making use of definitions (\ref{r}) and (\ref{kappa}), $C_{F}$ simply
reads:%
\begin{equation}
C_{F}\left(  z\right)  =\kappa,\qquad\forall z. \label{CFdef}%
\end{equation}

\subsubsection{The Laser Equations in the Uniform Field Limit}

Substitution of expressions (\ref{CP}), (\ref{CDdef}) and (\ref{CFdef}) into
Eqs. (\ref{dFdt}), (\ref{dPdt}) and (\ref{dDdt}) yields%
\begin{gather}
\frac{\partial F}{\partial t}+v\frac{\partial F}{\partial z}=\kappa\left(
P-F\right)  ,\label{dFU}\\
\frac{\partial P}{\partial t}=\gamma_{\bot}\left(  FD-P\right)  ,\label{dPU}\\
\frac{\partial D}{\partial t}=\gamma_{||}\left[  r-D-\tfrac{1}{2}\left(
FP^{\ast}+F^{\ast}P\right)  \right]  . \label{dDU}%
\end{gather}
We finally need to consider the boundary condition that applies to these
equations. By considering the definition (\ref{FPD}) for $F$%
\[
F\left(  0,t\right)  =\frac{\alpha^{\prime}\left(  0,t\right)  }%
{\alpha_{\mathrm{s}}\left(  0\right)  },\;F\left(  L_{\mathrm{m}},t\right)
=\frac{\alpha^{\prime}\left(  L_{\mathrm{m}},t\right)  }{\alpha_{\mathrm{s}%
}\left(  L_{\mathrm{m}}\right)  }.
\]
Making the quotient of both quantities we have%
\[
\frac{F\left(  0,t\right)  }{F\left(  L_{\mathrm{m}},t\right)  }=\frac
{\alpha_{\mathrm{s}}\left(  L_{\mathrm{m}}\right)  }{\alpha_{\mathrm{s}%
}\left(  0\right)  }\frac{\alpha^{\prime}\left(  0,t\right)  }{\alpha^{\prime
}\left(  L_{\mathrm{m}},t\right)  },
\]
which, making use of Eqs. (\ref{BCs}) and (\ref{BCprime}) yields%
\begin{equation}
F\left(  0,t\right)  =F\left(  L_{\mathrm{m}},t\right)  . \label{BCUFL}%
\end{equation}
We thus see that the boundary condition for the field amplitude is
\textit{periodic} (we note that this is not due to the uniform field limit but
to the very definition of $F$). This is of great importance as will allow us,
owing to the Fourier theorem, to decompose $F$ in terms of periodic functions.

Before studying the obtained Maxwell--Bloch equations (\ref{dFU}%
)--(\ref{dDU}), let us demonstrate that $F$, $P$, and $D$, are equivalent to
the original variables, apart from constant scale factors. From definitions
(\ref{FPD}), and using the uniform field limit results developed in the
previous section, we have%
\begin{align*}
F\left(  z,t\right)   &  =\frac{2}{\sqrt{\gamma_{||}\gamma_{\bot}}}%
\alpha^{\prime}\left(  z,t\right)  ,\\
P\left(  z,t\right)   &  =2i\frac{r}{d_{0}}\sqrt{\frac{\gamma_{\bot}}%
{\gamma_{||}}}\sigma_{21}^{\prime}\left(  z,t\right)  ,\\
D\left(  z,t\right)   &  =\frac{r}{d_{0}}d^{\prime}\left(  z,t\right)  .
\end{align*}
Thus $F$ has the meaning of a laser field amplitude, $P$ has the meaning of
material polarization and $D$ has the meaning of population difference.

Equations (\ref{dFU})--(\ref{dDU}) allow to study two types of laser
operation: singlemode and multimode. The method for deriving the laser
equations in the uniform field limit we have followed here was presented in
\cite{deValcarcel03} (see also \cite{nonresonant}), where an application to
multimode emission was addressed. In the following we shall concentrate on the
singlemode laser.

\section{The single--mode laser equations}

In the previous section we have derived the laser equations in the uniform
field limit for a resonant laser. For arbitrary detuning it can be
demonstrated that the laser equations in the uniform--field limit read
\cite{nonresonant}%
\begin{gather}
\frac{\partial F}{\partial t}+v\frac{\partial F}{\partial z}=\kappa\left(
P-F\right)  ,\label{ufl1}\\
\frac{\partial P}{\partial t}=\gamma_{\bot}\left[  FD-\left(  1+i\Delta
_{c}\right)  P\right]  ,\label{ufl2}\\
\frac{\partial D}{\partial t}=\gamma_{||}\left[  r-D-\tfrac{1}{2}\left(
FP^{\ast}+F^{\ast}P\right)  \right]  , \label{ufl3}%
\end{gather}
where $\Delta_{c}=\left(  \omega_{c}-\omega_{21}\right)  /\gamma_{\bot}$ is
the atom--cavity detuning parameter. These equations are complemented with the
periodic boundary condition
\begin{equation}
F\left(  0,t\right)  =F\left(  L_{\mathrm{m}},t\right)  .
\end{equation}

The fact that the boundary condition is periodic means that the field can be
written in the form
\begin{equation}
F\left(  z,t\right)  =\sum_{m=-\infty}^{+\infty}F_{m}\left(  t\right)
e^{iq_{m}z}, \label{modos}%
\end{equation}
where $q_{m}=m2\pi/L_{\mathrm{m}}$ with $m$ an integer. This means that the
intracavity field is, in general, a superposition of longitudinal modes of the
empty (i.e., without amplifying medium) cavity. In order to see this clearly,
let us solve Eq. (\ref{ufl1}) for the empty cavity and ignoring cavity losses,
i.e., Eq. (\ref{ufl1}) with its right--hand side equal to zero. Its solutions
have the form
\begin{equation}
F\left(  z,t\right)  =\sum_{m=-\infty}^{+\infty}F_{m}e^{i\left(  q_{m}%
z-\omega_{m}t\right)  },
\end{equation}
with $\omega_{m}=vq_{m}$. Now we must notice that the actual field is not
$F\left(  z,t\right)  $ but $F\left(  z,t-\tau\right)  $ with $\tau=z\Delta
t/L_{\mathrm{m}}$ and $\Delta t=\left(  L_{\mathrm{c}}-L_{\mathrm{m}}\right)
/c$, as we introduced new fields in Eqs. (\ref{cambio1}) and (\ref{cambio2})
(remind that the field $F$ is proportional to the field $\alpha^{\prime}$
which is different from the actual field $\alpha$). Then, the actual field (we
will not introduce a new symbol for it) is
\begin{align}
F\left(  z,t\right)   &  =\sum_{m=-\infty}^{+\infty}F_{m}\exp\left[  i\left(
q_{m}z-\omega_{m}z\frac{L_{\mathrm{c}}-L_{\mathrm{m}}}{cL_{\mathrm{m}}}%
-\omega_{m}t\right)  \right] \\
&  =\sum_{m=-\infty}^{+\infty}F_{m}\exp\left[  i\left(  k_{m}z-\omega
_{m}t\right)  \right]  ,
\end{align}
with%
\begin{equation}
k_{m}=q_{m}-\frac{\omega_{m}}{c}\frac{L_{\mathrm{c}}-L_{\mathrm{m}}%
}{L_{\mathrm{m}}}=m\frac{2\pi c}{L_{\mathrm{c}}}, \label{k}%
\end{equation}
where $\omega_{m}=vq_{m}$ has been used. Eq. (\ref{k}) shows clearly that the
actual field appears decomposed into empty--cavity modes.

Thus Eqs. (\ref{ufl1}--\ref{ufl3}) can describe multilongitudinal mode
emission when $\partial F/\partial z$ is non null and, when $\partial
F/\partial z=0$, this model can describe only singlemode emission. The
question now is: Should we keep the spatial derivative always? Or, in other
words, when will the laser emit in a single mode and when in several
longitudinal modes? In 1968 Risken and Nummedal \cite{RN} and, independently,
Graham and Haken \cite{GH}, demonstrated that Eqs. (\ref{ufl1}--\ref{ufl3})
predict the existence of multilongitudinal mode emission if certain conditions
are verified. We are not going to treat the Risken--Nummedal--Graham--Haken
instability here (see, e.g.,
\cite{NarducciAbraham,WeissVilaseca,Khanin,Mandel} or \cite{nonresonant} for a
recent review), it will suffice to say that for multilongitudinal mode
emission to occur the two necessary conditions are: (i), a large enough pump
value (in resonance, $\Delta=0$, $r$ must be larger than nine and remember
that the laser threshold in these conditions, given Eq. (\ref{threshold}),
equals unity; out of resonance even more pump is required), and most
importantly; (ii), the cavity length must be large, unrealistically large for
common lasers. Then for short enough cavities (and this is not a restrictive
condition at all for most lasers) the laser will emit in a single longitudinal
mode. In this case, the spatial derivative in Eq. (\ref{ufl1}) can be removed
and we are left with the Maxwell--Bloch equations for a singlemode laser. We
must insist that all this is true for homogeneously broadened lasers and
cannot be applied to inhomogeneously broadened ones, see \cite{nonresonant}.

Then for singlemode lasers we can take $\partial F/\partial z=0$. It is
particularly interesting to write down the singlemode laser equations in
resonance ($\Delta_{c}=0$). Let us write the field and atomic polarization in
the following way%
\begin{equation}
F=Ee^{i\phi},\ \ \ P=\left(  P_{re}+iP_{im}\right)  e^{i\phi},
\end{equation}
with $E$ a real quantity. Now Eqs. (\ref{ufl1},\ref{ufl2}), with $\Delta
_{c}=0$, read
\begin{gather}
\dot{E}=\kappa\left(  P_{re}-E\right)  ,\\
\dot{\phi}=\kappa\frac{P_{im}}{E},\\
\dot{P}_{re}=\gamma_{\bot}\left(  ED-P_{re}\right)  +\dot{\phi}P_{im},\\
\dot{P}_{im}=-\gamma_{\bot}P_{im}-\dot{\phi}P_{re},
\end{gather}
where the dot means total derivative with respect to time. By suitably
combining the first, the second and last equations one obtains
\begin{equation}
\frac{\dot{P}_{im}}{P_{im}}+\frac{\dot{E}}{E}=-\left(  \kappa+\gamma_{\bot
}\right)  ,
\end{equation}
from which%
\begin{equation}
P_{im}\left(  t\right)  =\frac{P_{im}\left(  0\right)  }{E\left(  0\right)
}E\left(  t\right)  e^{-\left(  \kappa+\gamma_{\bot}\right)  t}\rightarrow0,
\end{equation}
and then $\dot{\phi}\rightarrow0$ also. Thus, in resonance, the singlemode
laser equations reduce to only three real equations
\begin{gather}
\dot{E}=\kappa\left(  P-E\right)  ,\label{L1}\\
\dot{P}=\gamma_{\bot}\left(  ED-P\right)  ,\label{L2}\\
\dot{D}=\gamma_{||}\left(  r-D-EP\right)  , \label{L3}%
\end{gather}
where $P=P_{re}$.

The above set of equations is usually known as Haken--Lorenz equations. The
reason for this name is the following: Let us define the adimensional time
$\tau=\gamma_{\bot}t$, and the new variables and normalized relaxation rates
\begin{align}
X  &  =E,\ \ Y=P,\ \ Z=r-D,\\
\sigma &  =\frac{\kappa}{\gamma_{\bot}},\ \ \ b=\frac{\gamma_{||}}%
{\gamma_{\bot}}.
\end{align}
These new variables verify
\begin{gather}
\frac{\mathrm{d}}{\mathrm{d}\tau}X=\sigma\left(  Y-X\right)  ,\\
\frac{\mathrm{d}}{\mathrm{d}\tau}Y=rX-Y+XZ,\\
\frac{\mathrm{d}}{\mathrm{d}\tau}Z=b\left(  Z-XY\right)  .
\end{gather}
These are the Lorenz equations \cite{Lorenz}, which are a very simplified
model proposed by Edward N. Lorenz in 1961 for the baroclinic instability, a
very schematic model for the atmosphere. They constitute a paradigm for the
study of deterministic chaos as they constitute the first model that was
found, by Lorenz himself, to exhibit deterministic chaos. It was Herman Haken
who, in 1975 \cite{Haken75}, demonstrated the astonishing isomorphism existing
between the Lorenz model and the resonant laser model that we have just
demonstrated. After this recognition, the study of deterministic chaos in
lasers became a very active area of research (see, e.g.,
\cite{Haken,NarducciAbraham,WeissVilaseca,Khanin,Mandel}).

The Haken--Lorenz model exhibits periodic and chaotic solutions, and several
routes to chaos can be found in its dynamics. The equations can be numerically
integrated easily, e.g. with Mathematica, and we refer the interest reader to
\cite{Haken,NarducciAbraham,WeissVilaseca,Khanin,Mandel} for suitable
introductions into these fascinating subjects. Here we shall only briefly
comment on a particular point.

Eqs. (\ref{L1},\ref{L3}) have two sets of stationary solutions: The laser off
solution ($E=P=0$ and $D=r$), and the lasing solution ($E=P=\pm\sqrt{r-1}$ and
$D=1$) that exists for $r\geq1$. A linear stability analysis of this last
solution shows that it becomes unstable when $\kappa\geq\gamma_{\bot}%
+\gamma_{||}$ (this condition is know as "bad cavity" condition) and $r\geq
r_{HB}$ with%
\begin{equation}
r_{HB}=\frac{\kappa\left(  \kappa+3\gamma_{\bot}+\gamma_{||}\right)  }%
{\kappa-\left(  \gamma_{\bot}+\gamma_{||}\right)  }, \label{Hopf}%
\end{equation}
whose minimum value is $9$ for $\gamma_{||}=0$ and $\kappa=3\gamma_{\bot}$.
Notice that as we are considering the resonant case for which the lasing
threshold is $r_{on}=1$, $r_{HB}=9$ means that the adimensional effective pump
$r$ must be, at least, nine time above the instability threshold.

From Eq. (\ref{Hopf}) we can say that the singlemode solution is always stable
for good cavities ($\kappa<\gamma_{\bot}+\gamma_{||}$) and also for bad
cavities if the pump is small ($r<r_{HB}$), but for bad cavities and large
pump, the stationary solution becomes unstable (through a Hopf bifurcation)
and a self--pulsing occurs (e.g., chaotic oscillations). The condition $r\geq
r_{HB}$ (remember that $r_{HB}\geq9$) is usually considered as a very
restrictive condition (we insist, the laser should be pumped nine times above
threshold, and this is quite a large pump value!) but we have seen that we
must be careful when interpreting the pump parameter $r$. In fact, if one
considers a three--level laser and uses Eqs. (\ref{r}), (\ref{kappa}),
(\ref{gammapar3L}) and (\ref{pump3L}), one can write Eq. (\ref{Hopf}) in terms
of the actual pump strength and decay rate and gets that the instability
threshold to lasing threshold pumps ratio can be very close to unity.

We can show this easily. Let us recall equation (\ref{r}), that relates the
adimensional effective pump parameter $r$ with the inversion in the absence of
fields $d_{0}$%
\begin{align}
r  &  =Gd_{0}\\
G  &  \equiv\frac{2gL_{\mathrm{m}}}{\gamma_{\bot}c\left\vert \ln
\mathcal{R}^{2}\right\vert }, \label{G}%
\end{align}
as well as Eq. (\ref{pump3L}) that relates $d_{0}$ to the actual physical pump
parameter $R$ in three--level lasers. By taking into account that $r_{on}=1$
and $r_{HB}=9$ it is easy to see that%
\begin{equation}
\frac{R_{HB}^{3L}}{R_{on}^{3L}}=\frac{\left(  G+9\right)  \left(  G-1\right)
}{\left(  G+1\right)  \left(  G-9\right)  },
\end{equation}
which for large $G$ simplifies to
\begin{equation}
\frac{R_{HB}^{3L}}{R_{on}^{3L}}\approx1+\frac{16}{G}+\mathcal{O}\left(
G^{-2}\right)  ,
\end{equation}
i.e., for three--level lasers the "very restrictive" condition $r_{HB}=9$
turns out to be an easy condition in terms of pumping ($R_{HB}^{3L}%
/R_{on}^{3L}=1+\varepsilon$) when the gain parameter $G$, Eq. (\ref{G}), is
large enough.

\section{Conclusion}

In this article we have presented a self--contained derivation of the
semiclassical laser equations. We have paid particular attention to: (i)\ the
adequacy of the standard two--level model to more realistic three-- and
four--level systems; and (ii), the derivation of the laser equations in the
uniform field limit. We think that our presentation could be useful for a
relatively rapid, as well as reasonably rigorous, introduction of the standard
laser theory. This should be complemented with a detailed analysis of the
stability of the stationary laser solution (see, e.g.,
\cite{Haken,NarducciAbraham,WeissVilaseca,Khanin,Mandel}) which we do not
treat here for the sake of brevity.\bigskip

{\LARGE Acknowledgements}

This work has been financially supported by the Spanish Ministerio de Ciencia
y Tecnolog\'{\i}a and European Union FEDER (Project
FIS2005-07931-C03-01).\bigskip

\section{Appendix}

In this Appendix we demonstrate that the adiabatic elimination of the atomic
coherence in Eqs. (\ref{B1},\ref{B2}) consists in making $\partial_{t}%
\sigma_{12}=0$.

Consider the evolution equation%
\begin{equation}
\frac{\mathrm{d}}{\mathrm{d}t}f\left(  t\right)  =-\gamma f\left(  t\right)
+g\left(  t\right)  , \label{f}%
\end{equation}
which must be complemented with the equation of evolution of $g\left(
t\right)  $. Notice that Eq. (\ref{f}) coincides with Eq. (\ref{B2}) for
$f\left(  t\right)  =\sigma_{12}\exp\left(  i\delta t\right)  $, $g\left(
t\right)  =id\alpha^{\ast}\exp\left(  i\delta t\right)  $, and $\gamma
=\gamma_{\bot}$. Now we define the new variable $\bar{f}=f\exp\left(  -\gamma
t\right)  $ that verifies%
\begin{equation}
\frac{\mathrm{d}}{\mathrm{d}t}\bar{f}=e^{-\gamma t}g\left(  t\right)  ,
\end{equation}
from which%
\begin{equation}
\bar{f}\left(  t\right)  =\bar{f}\left(  0\right)  +\int_{0}^{t}dt^{\prime
}e^{-\gamma t^{\prime}}g\left(  t^{\prime}\right)  ,
\end{equation}
and then%
\begin{equation}
f\left(  t\right)  =f\left(  0\right)  e^{-\gamma t}+\int_{0}^{t}dt^{\prime
}e^{-\gamma\left(  t-t^{\prime}\right)  }g\left(  t^{\prime}\right)  .
\end{equation}
Integrating by parts and ignoring the first (decaying) term%
\begin{equation}
f\left(  t\right)  =\frac{1}{\gamma}\left[  g\left(  t\right)  -\int_{0}%
^{t}dt^{\prime}e^{-\gamma\left(  t-t^{\prime}\right)  }\frac{\mathrm{d}%
}{\mathrm{d}t^{\prime}}g\left(  t^{\prime}\right)  \right]  .
\end{equation}
Then, after repeatedly integrating by parts, one finally obtains%
\begin{equation}
f\left(  t\right)  =\frac{1}{\gamma}\left[  1-\frac{1}{\gamma}\frac
{\mathrm{d}}{\mathrm{d}t}+\frac{1}{\gamma^{2}}\frac{\mathrm{d}^{2}}%
{\mathrm{d}t^{2}}-\ldots\right]  g\left(  t\right)  ,
\end{equation}
and thus for large enough $\gamma$ one can approximate $f\left(
t\right) \approx\gamma^{^{\prime}-1}g\left(  t\right)  $, which is
the result one obtains by making
$\frac{\mathrm{d}}{\mathrm{d}t}f\left(  t\right)  =0$ in Eq.
(\ref{f}), as we wanted to demonstrate.

\end{document}